\documentclass[preprint,12pt]{elsarticle}

\usepackage{amssymb}

\usepackage[nodots]{numcompress}

\usepackage{hyperref}

\usepackage{sudretstyle}
\usepackage{amsmath}
\usepackage{amssymb}
\usepackage{amsfonts}
\usepackage{amsbsy}
\usepackage{bbm}
\usepackage[FIGTOPCAP]{subfigure}
\usepackage{array} 
\usepackage{booktabs} 
\usepackage{multirow} 

\usepackage{color}

\journal{Structural Safety}

\begin{document}

\begin{frontmatter}



\title{Assessment of the lognormality assumption of seismic fragility curves using non-parametric representations}


\author[myaddress]{B. Sudret}
\ead[url]{www.ibk.ethz.ch/su/people/sudretb/index\_EN}

\author[myaddress]{C. Mai}

\author[myaddress]{K. Konakli \corref{mycorrespondingauthor}}
\cortext[mycorrespondingauthor]{Corresponding author}
\ead{konakli@ibk.baug.ethz.ch}

\address[myaddress]{ETH Z\"{u}rich, Institute of Structural Engineering, Chair of Risk, Safety \& Uncertainty Quantification, Stefano-Franscini-Platz 5, CH-8093 Z\"{u}rich, Switzerland}

\begin{abstract}
Fragility curves are commonly used in civil engineering to estimate the vulnerability of structures to 
earthquakes. The probability of failure associated with a prescribed criterion (e.g. the maximal inter-storey 
drift ratio exceeding a prescribed threshold) is represented as a function of the intensity of the earthquake ground motion
(e.g. peak ground acceleration or spectral acceleration). The classical approach consists in assuming a 
lognormal shape of the fragility curves. In this paper, we introduce two non-parametric approaches to 
establish the fragility curves without making any assumption, namely the binned Monte Carlo simulation approach and kernel density estimation. As an illustration, we compute the fragility curves of a 3-storey steel
structure, accounting for the nonlinear behavior of the system. The curves obtained with the proposed approaches 
are compared with each other and with those obtained using the classical lognormal assumption. It is shown that the lognormal curves differ significantly from their non-parametric counterparts.
\end{abstract}


\end{frontmatter}


\section{Introduction}

The severe socio-economic consequences of several recent earthquakes highlight the need for proper seismic risk assessment as a basis for efficient decision making on mitigation actions and disaster planning. To this end, the probabilistic performance-based earthquake engineering (PBEE) framework has been developed, which allows explicit evaluation of performance measures that serve as decision variables (DV) (\eg monetary losses, casualties, downtime) accounting for all prevailing uncertainties (\eg characteristics of ground motions, structural properties, damage occurrence). The key steps in the PBEE framework comprise the identification of seismic hazard, evaluation of structural response, damage analysis and eventually, consequence evaluation. In particular, the mean annual frequency of exceedance of a DV is evaluated as \cite{Porter2003, Baker2008a, Gunay2013}:
\begin{equation}
 \lambda(DV)=\int\int\int P(DV|DM) \, \di P(DM|EDP) \, \di P(EDP|IM) \, \di \lambda(IM)
 \label{eqPBEE}
\end{equation}
in which $P(x|y)$  is the conditional probability of $x$ given $y$, $DM$  is a damage measure typically defined according to repair costs (\eg light, moderate or severe damage), $EDP$ is an engineering demand parameter obtained from structural analysis (\eg force, displacement, drift ratio), $IM$ is an intensity measure characterizing the ground motion severity (\eg peak ground acceleration, spectral acceleration) and $\lambda(IM)$ is the annual frequency of exceedance of the $IM$. Determination of the probabilistic model $P(EDP|IM)$ constitutes a major challenge in the PBEE framework since the earthquake excitation contributes the most significant part to the uncertainty in the $DV$. This step of the analysis is the focus of the present paper.

The conditional probability $P(EDP \geq \overline{edp}|IM)$, where $\overline{edp}$ denotes an acceptable demand threshold, is commonly represented graphically in the shape of the so-called demand fragility curves \cite{Mackie2005}. Thus, a demand fragility curve represents the probability that an engineering demand parameter exceeds a prescribed threshold as a function of an intensity measure of the earthquake motion. For the sake of simplicity, demand fragility curves are simply denoted fragility curves in the following analysis, which is also typical in the literature see \eg \cite{Ellingwood2009, Seo2012}. 
We note however that in other publications the term \emph{ fragility} is also used for $P(DM\geq\overline{dm}|IM)$ and $P(DM\geq\overline{dm}|EDP)$, \ie the conditional probability of damage exceeding a threshold $\overline{dm}$ given the intensity measure \cite{Banerjee2007} or the engineering demand parameter \cite{Baker2008a,Gunay2013}, respectively.

Originally introduced in the early 1980's for nuclear safety evaluation \cite{Richardson1980}, fragility 
curves are nowadays widely used for multiple purposes, \eg seismic loss estimation \cite{Pei2009}, estimation 
of the collapse risk of structures in seismic regions \cite{Eads2013}, design 
checking process \cite{Dukes2012}, evaluation of the effectiveness of retrofit 
measures \cite{Guneyisi2008}, etc. It is noteworthy that novel methodological contributions to fragility analysis have been made in recent years, including the development of multi-variate fragility functions \cite{Seyedi2010}, incorporation of Bayesian updating \cite{Gardoni2002a} and time-dependent fragility curves \cite{Ghosh2010}. However, the classical fragility curves remain a popular tool in seismic risk assessment and recent literature is rich with applications on various type of structures, such as irregular buildings \cite{Seo2012}, underground tunnels \cite{Argyroudis2012}, a pile-supported wharf \cite{Chiou2011}, wind turbines \cite{Quilligan2012}, nuclear power plant equipments \cite{Borgonovo2013}. The estimation of such curves is the focus of this paper.

Fragility curves are typically classified into four categories according to the data
sources, namely analytical, empirical, judgment-based or hybrid fragility curves \cite{Rossetto2005}.
Analytical fragility curves are derived from data obtained by analyses of structural models.
Empirical fragility curves are based on the observation of earthquake-induced damage reported in 
post-earthquake surveys. Judgment-based curves are estimated by expert panels specialized
in the field of earthquake engineering. Hybrid curves are typically obtained 
by combining data from different sources. Each category of fragility curves 
has its own advantages as well as drawbacks.
In this paper, analytical fragility curves established 
using data collected from numerical structural analyses are of interest.

The typical approach to compute analytical fragility curves presumes that the curves have 
the shape of a lognormal cumulative distribution function \cite{Shinozuka2000b,Ellingwood2001}.
This approach is therefore considered \emph{parametric}.
The parameters of the lognormal distribution are determined either by maximum likelihood 
estimation  \cite{Shinozuka2000b,Zentner2010a,Seyedi2010} or by fitting a linear 
probabilistic seismic demand model in the log-scale \cite{Ellingwood2009,Gencturk2008, Jeong2012, Banerjee2008}.
The assumption of lognormal fragility curves is
almost unanimous in the literature due to the computational convenience as well as due to the facility for 
combining such curves with other elements of the seismic probabilistic risk assessment framework. However, the validity of such assumption remains questionable.

In the present work, we present two novel non-parametric approaches to establish the fragility curves without making any assumption, namely the binned Monte Carlo simulation (bMCS) and kernel density estimation (KDE). The main advantage of bMCS over traditional Monce Carlo simulation approaches (\cite{Hwang1994, Lupoi2005}) is that it avoids the bias induced by scaling ground motions to predefined intensity levels. In the KDE approach, we introduce a statistical methodology for fragility estimation, which also opens new paths for estimation of multi-dimensional fragility functions. The proposed methods are subsequently used to investigate the validity of the lognormal assumption in a case study where we develop fragility curves for different thresholds of the maximum inter-storey drift ratio of an example building structure. The proposed methodology can be applied in a straightforward manner to other types of structures or classes of structures or using different failure criteria.

The computation of fragility curves requires a sufficiently large number of transient dynamic analysis of the structure 
under seismic excitations that are either recorded or synthetic. Due to the lack of recorded signals 
with the properties of interest (\eg earthquake magnitude, duration, etc.), it is common practice to 
generate suitable samples of synthetic ground motions \cite{Choi2004,Guneyisi2008}. 

The paper is organized as follows: in Section 2, the method recently proposed by \cite{Rezaeian2008} to generate
synthetic earthquakes, used in the example in Section 4, is briefly recalled. This method is selected because it allows to account for uncertainty in the intensity level of the motions for a given earthquake scenario. The different approaches 
for establishing the fragility curves, namely the classical lognormal, bMCS and KDE approaches, are 
presented in Section 3. In Section 4, we compute the fragility curves of 
a steel frame structure subject to seismic excitations using the aforementioned approaches and compare the results.

\vspace{-6pt}
\section{Recorded and synthetic ground motions}
\vspace{-2pt}
\subsection{Recorded ground motions: notation}
\label{sec2.1}
Let us consider a recorded earthquake accelerogram $a(t)$, $t\in [0,T]$ where $T$ is the total duration of the 
motion. The peak ground acceleration reads: $PGA= \max\limits_{\substack{t\in[0,T]}} \, \lvert a(t) \lvert $. The 
Arias intensity $I_a$ is defined by:
\begin{equation}
I_a=\frac{\pi}{2g} \int\limits_0^T {a^2(t)}\, \mathrm{d} t
\label{eq:Ia}
\end{equation}
Defining the cumulative square acceleration by:
\begin{equation}
I(t)=\frac{\pi}{2g} \int\limits_0^t \,{a^2(\tau)}\, \mathrm{d} \tau \, ,
\label{eq:It}
\end{equation}
one defines the time instant $t_{\alpha}$ by:
\begin{equation}
 t_{\alpha}: \qquad I(t_{\alpha})=\alpha I_a \qquad \alpha \in [0,1]
\end{equation}
Remarkable properties of a recorded seismic motion are the duration of the \emph{strong motion phase} $D_{5-95}=t_{95\%} - t_{5\%}$
and the instant at the middle of the strong-shaking phase $t_{mid} \equiv t_{45\%}$.
\subsection{Simulation of synthetic ground motions}

For the sake of completeness, we summarize in this section the parameterized approach proposed by \cite{Rezaeian2008} in order to simulate synthetic ground
motions. The seismic acceleration $a(t)$ is represented as a non-stationary
process. Der Kiureghian and Rezaeian separate the non-stationarity into
two components, namely a spectral and a temporal one, by means of a
modulated filtered Gaussian white noise:
\begin{equation} 
  a(t)=  \frac{q(t,\ve{\alpha}) }{\sigma_h{(t)}}
    \int\limits_{0}^t \,{h \bra{ t-\tau,\ve{\lambda}\prt{\tau} }
      \omega(\tau)}\, \mathrm{d} \tau 
\label{eq:at}
\end{equation} in which $q(t,\ve{\alpha})$ is the deterministic
non-negative \emph{modulating function}, the integral is the non-stationary response of a 
linear filter subject to a Gaussian white noise excitation and $\sigma_h{(t)}$ is the standard deviation
of the response process.
The Gaussian white-noise process denoted by $\omega(\tau)$ will pass through a
filter $h \bra{t-\tau,\ve{\lambda}(\tau)}$ which is selected as the pseudo-acceleration response of a 
single-degree-of-freedom (SDOF) linear oscillator:
\begin{equation}
  \begin{aligned}
   & h \bra{t-\tau,\ve{\lambda}(\tau)}=0 \quad \text{for} \quad t < \tau \\ 
    & h \bra{t-\tau,\ve{\lambda}(\tau)}=
    \frac{\omega_f(\tau)}{\sqrt{1-\zeta_f^2(\tau)}} \mathrm{exp}
    \bra{-\zeta_f(\tau) \omega_f(\tau) (t-\tau) }  \sin \bra{
      \omega_f(\tau) \sqrt{1-\zeta_f^2(\tau)} (t-\tau) } \\
      & \quad \quad \quad \quad \quad \quad \quad \quad \text{for}
    \quad t \geq \tau
  \end{aligned}
\end{equation} 
where $\ve{\lambda}(\tau)=\prt{ \omega_f(\tau),\zeta_f(\tau) }$ is the
vector of time-varying parameters of the filter $h$. Note that $\omega_f(\tau)$
and $\zeta_f(\tau)$ are the filter's natural frequency and damping ratio
at instant $\tau$, respectively. They are related to the evolving predominant
frequency and bandwidth of the ground motion that is to be represented. The statistical analysis
of real signals shows that $\zeta_f(\tau)$ may be taken as a
constant $\prt{ \zeta_f(\tau) \equiv \zeta }$ while the predominant
frequency varies approximately linearly in time \cite{Rezaeian2010}:
\begin{equation} 
  \omega_f(\tau) =\omega_{mid} + \omega'(\tau- t_{mid})
  \label{eq3}
\end{equation} 
In \eqrefe{eq3} $t_{mid}$ is the instant at which 45\% of the Arias
intensity $I_a$ is reached, $\omega_{mid}=\omega_f(t_{mid})$ is the filter's frequency at
instant $t_{mid}$ and $\omega'$ is the slope of the linear evolution. After
being normalized by the standard deviation $\sigma_h{(t)}$, the integral
in \eqrefe{eq:at} becomes a unit variance process with time-varying
frequency and constant bandwidth.
The non-stationarity in intensity is then captured by the modulating
function $q(t,\ve{\alpha})$. This time-modulating function determines
the shape, intensity and duration $T$ of the signal. A Gamma-like
function is typically used \cite{Rezaeian2010}:
\begin{equation}
 q(t,\ve{\alpha})=\alpha_1 t^{\alpha_2-1}\mathrm{exp}(-\alpha_3t)
\end{equation}
where $\ve{\alpha}=\acc{\alpha_1,\alpha_2,\alpha_3}$ is
directly related to the energy content of the signal through the quantities $I_a$, $D_{5-95}$ 
and $t_{mid}$ defined in Section \ref{sec2.1}, see \cite{Rezaeian2010} for details.
For computational purposes, the acceleration in \eqrefe{eq:at} can be discretized as follows:
\begin{equation} 
  \hat{a}(t)=q(t,\ve{\alpha})\sum_{i=1}^n s_i \prt{
    t,\ve{\lambda}(t_i)} \, U_i
  \label{eq:at2}
\end{equation} where the standard normal random variable $U_i$
represents an impulse at instant $t_i=i \times \dfrac{T}{n} \, , \, i=1
\enu n $, ($T$ is the total duration) and $s_i(t,\ve{\lambda}(t_i))$ is
given by:
\begin{equation}
  s_i(t,\ve{\lambda}(t_i)) = \dfrac{h\bra{t-t_i,\ve{\lambda}(t_i)} }{\sqrt{\sum_{j=1}^i h^2 \bra{t-t_j,\ve{\lambda}(t_j)} }}
\end{equation}
As a summary, the considered seismic generation model consists of three temporal parameters
$\left(\alpha_1,\alpha_2,\alpha_3\right)$, three spectral parameters
$\left(\omega_{mid}, \omega', \zeta_f\right) $ and the standard Gaussian
random vector $\ve{U}$ of size $n$. 
Note that the full model proposed by \cite{Rezaeian2010} includes some additional parameters that are related to earthquake and
site characteristics, \eg the type of faulting of the earthquake (strike-slip fault or reverse fault), the closest distance
from the recording site to the ruptured area and the shear-wave velocity of the top 30~m of the site soil. 
A methodology for determining the temporal and spectral parameters according to earthquake and site characteristics is described in \cite{Rezaeian2010}. For the sake of simplicity, in this paper these parameters are directly generated from appropriate statistical models.

\section{Computation of fragility curves}
\label{sec3}
Fragility curves represent the probability of failure of the system associated with a specified criterion for a given intensity measure ($IM$) of the earthquake motion. Herein probabilities of failure represent probabilities of exceeding demand limit states. The engineering demand parameter typically used for buildings is 
the maximal inter-storey drift ratio $\Delta$, \ie the maximal difference of horizontal 
displacements between consecutive storeys normalized by the storey height \cite{Seo2012}. Thus 
the fragility curve is cast as follows:
\begin{equation}
 \text{Frag}(IM;\,\delta_o)=\mathbb{P}[\Delta\geq \delta_o \lvert IM]
 \label{eq7}
\end{equation}
in which $\text{Frag}(IM;\,\delta_o)$ denotes the fragility at the given $IM$ for a prescribed demand threshold $\delta_o$ of the inter-storey drift ratio. In order to establish the fragility curves, a number $N$ of 
transient finite element analyses of the structure under consideration are used to provide paired values $\acc{\prt{IM_i,\Delta_i}, i=1 \enu N}$.
\subsection{Classical approach}
\label{sec3.1}
The classical approach to establish fragility curves consists in
assuming a \emph{lognormal shape} for the curves in \eqrefe{eq7}.
Two techniques are commonly used to estimate the parameters of the lognormal 
fragility curves, namely maximum likelihood estimation (MLE) and linear regression.

\subsubsection{Maximum likelihood estimation:}

One assumes that the fragility curves can be written in the following general form:
\begin{equation}
  \widehat{\text{Frag}}(IM;\,\delta_o) = \Phi \prt{\dfrac{\log IM - \log \alpha}{\beta}}
\label{eq9b}
\end{equation}
where $\Phi\left(t\right)= \int\limits_{-\infty}^t \, {e^{-u^2/2}}/{\sqrt{2\pi}} \mathrm{d} u$ is the standard Gaussian
cumulative distribution function (CDF), $\alpha$ is the ``median'' and $\beta$ is the ``log-standard 
deviation'' of the lognormal curve.
\cite{Shinozuka2000b} proposed 
the use of maximum likelihood estimation to determine these parameters as follows: 
Denote by $\omega$ the event that the demand threshold $\delta_o$ is reached or exceeded. Assume that $Y(\omega)$ is a random variable with Bernoulli distribution,
\ie $Y$ takes the value 1 (resp. 0) with probability $\text{Frag}(\cdot;\,\delta_o)$ 
(resp. $1-\text{Frag}(\cdot;\,\delta_o)$).
Considering a set of $i=1 \enum N$ ground motions, the likelihood function reads:
\begin{equation}
 \cl \prt{\alpha,\,\beta,\,\acc{IM_i,\,i=1 \enum N}} = \prod_{i=1}^N \bra{\text{Frag}(IM_i;\,\delta_o)}^{y_i} \, \bra{1- \text{Frag}(IM_i;\,\delta_o)}^{1- y_i} 
 \label{eq9c}
\end{equation}
where $IM_i$ is the intensity measure of the $i^{th}$ seismic motion and $y_i$ represents a realization
of the Bernoulli random variable $Y$, which takes the value 1 or 0 depending 
on whether the structure under the $i^{th}$ ground motion sustains the demand threshold $\delta_o$ or not.
The parameters $(\alpha,\, \beta)$ are obtained by maximizing the likelihood function. 
In practice, a straightforward optimization algorithm is applied on the log-likelihood function, \ie:
\begin{equation}
 \acc{\alpha^{\ast};\,\beta^{\ast}}\tr = \arg\max  \log \cl \prt{\alpha,\,\beta,\,\acc{IM_i,\,i=1 \enum N}}
 \label{eq9e}
\end{equation}

\subsubsection{Linear regression:}

One first assumes a \textit{probabilistic seismic demand model} which relates a structural
response quantity to an intensity measure of the earthquake motion.
More specifically, the
maximal inter-storey drift $\Delta$ is modelled by a lognormal
distribution whose log-mean value is a \emph{linear function}
of $\log IM$:
\begin{equation}
 \log \Delta = A \, \log IM + B + \zeta \,Z 
 \label{eq8}
\end{equation}
where $Z \sim \cn(0,1)$ is a standard normal variable.
Parameters $A$ and $B$ are determined by means of ordinary least squares estimation in a
log-log plot. This approach is
widely applied in the literature, see \eg \cite{ Ellingwood2001,Choi2004, Padgett2008,Zareian2007} among others. 
The procedure is based on the assumed linear relationship between $IM$ and $\Delta$ in 
the log-scale. Let us denote by $e_i$ the residual between the actual value $\log \Delta$
and the value predicted by the linear model: $e_i= \log \Delta_i - A \log
\prt{IM_i} - B$. Parameter $\zeta$ is obtained by:
\begin{equation}
 \zeta^2= \sum_{i=1}^{N} e_i^2 /\prt{N-2}
\end{equation}
\eqrefe{eq7} rewrites:
\begin{equation}
\begin{aligned}
 \widehat{\text{Frag}}(IM;\,\delta_o) &= \mathbb{P} \left[ \log \Delta  \geq \log \delta_o \right] 
 = 1- \mathbb{P} \left[ \log \Delta  \leq \log \delta_o \right]  \\
 & = \Phi \prt{\dfrac{\log IM - \prt{\log \delta_o - B}/A}{\zeta/A}}
\end{aligned}
\label{eq9}
\end{equation}
The median and log-standard deviation of the lognormal fragility curve in \eqrefe{eq9} are $\alpha=\expo{\prt{\log \delta_o - B}/A}$
and $\beta= \zeta/A$ respectively.

The above so-called classical approach is \emph{parametric} because it 
imposes the shape of the fragility curves in \eqrefe{eq9b} and \eqrefe{eq9} which is similar to a lognormal CDF when considered as a 
function of $IM$. 
In the sequel, we propose two \emph{non-parametric} approaches to compute fragility 
curves \emph{without} making such an assumption.
\subsection{Binned Monte Carlo simulation}
\label{sec3.2}
\begin{figure}[!ht]
\centering
\subfigure
	{
		\includegraphics[width=0.45\textwidth]{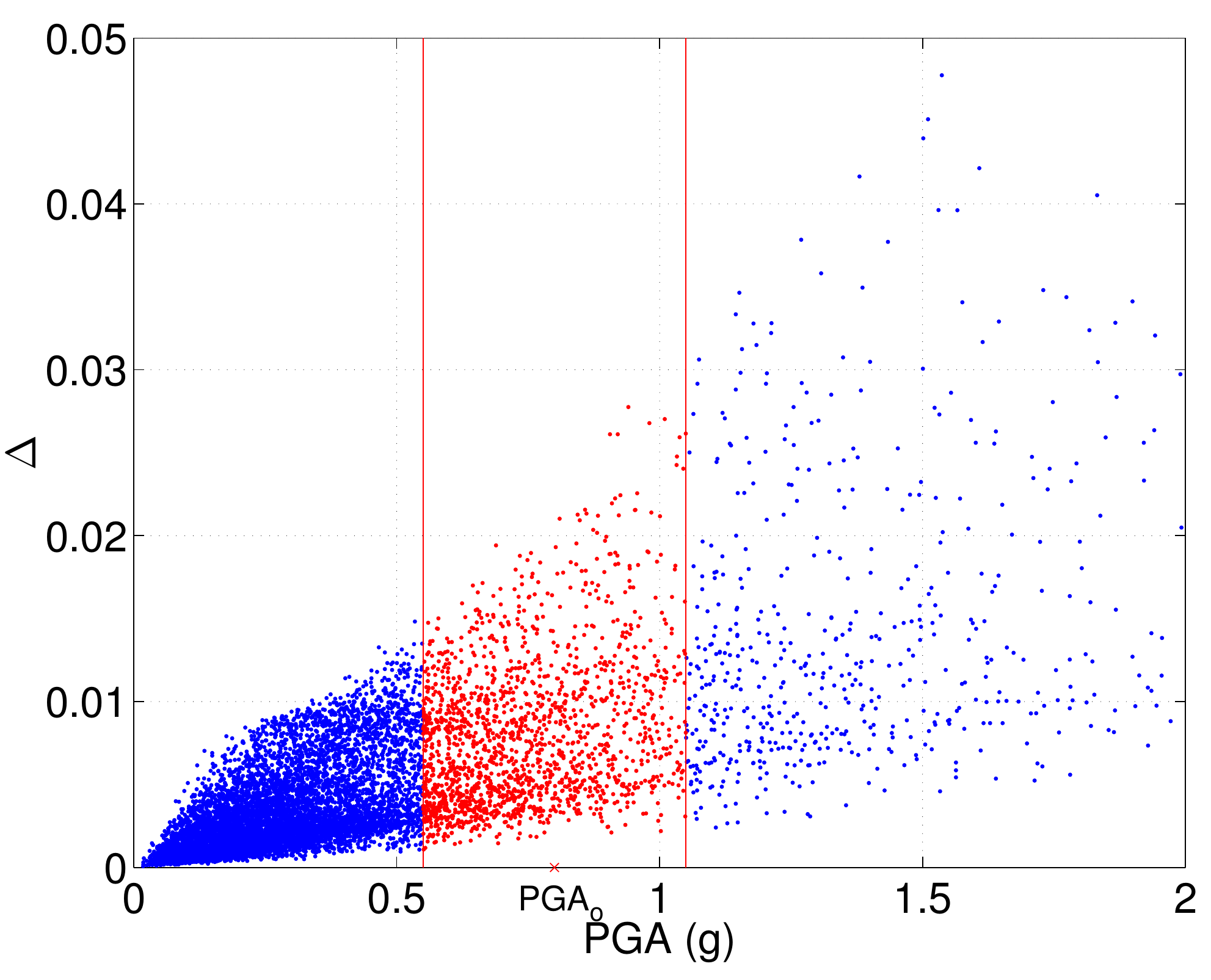}
	}
\subfigure
	{
		\includegraphics[width=0.45\textwidth]{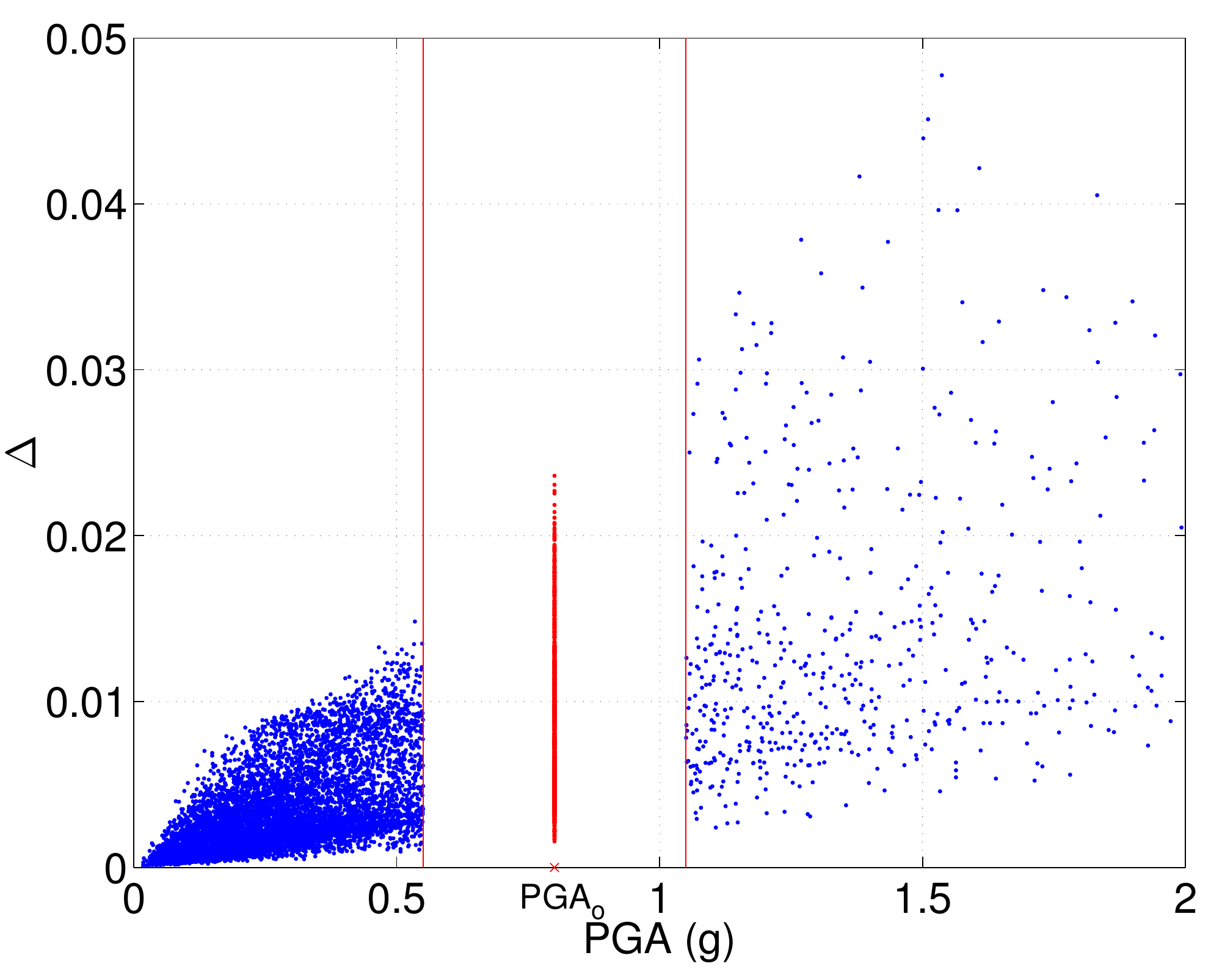}
	}
\caption{Scaling of ground motions and corresponding responses in the binned Monte Carlo simulation 
approach (the bin is enlarged for the sake of clarity).}
\label{fig1}
\end{figure}

Having at hand a large sample set of pairs $\acc{\prt{IM_j,\Delta_j},\,j=1 \enu N }$, it is 
possible to use a \emph{binned Monte Carlo simulation} (bMCS) to compute the fragility
curves, as described in the following. Let us consider a given abscissa $IM_o$. Within a small bin
surrounding $IM_o$, say $\bra{IM_o -h, IM_o +h}$ (see \figref{fig1}) one assumes that the
maximal drift $\Delta$ is linearly related to the $IM$. Note that this
assumption is exact in the case of linear structures but would only be
an approximation in the nonlinear case. Therefore, the maximal drift
$\Delta_j$ related to $IM_j \in \bra{IM_o -h, IM_o +h}$ is converted
into the drift $\Delta_j(IM_o)$ which is related to a similar input signal
having an intensity measure of $IM_o$ as follows:
\begin{equation}
 \Delta_j(IM_o)=\Delta_j \dfrac{IM_o}{IM_j}
 \label{eq13}
\end{equation}
The fragility curve at $IM_o$ is obtained by a
crude Monte Carlo estimator:
\begin{equation}
 \widehat{\text{Frag}} (IM_o)=\frac{N_f\prt{IM_o}}{N_s\prt{IM_o}}
\end{equation}
where $N_f\prt{IM_o}$ is the number of points in the bin such that $\Delta_j(IM_o) \geqslant \delta_o$ and $N_s(IM_o)$ is
the total number of points that fall into the bin $\bra{IM_o -h,IM_o +h}$.

\begin{figure}[!ht]
\centering
\subfigure
[Peak ground acceleration]
	{
		\includegraphics[width=0.45\textwidth]{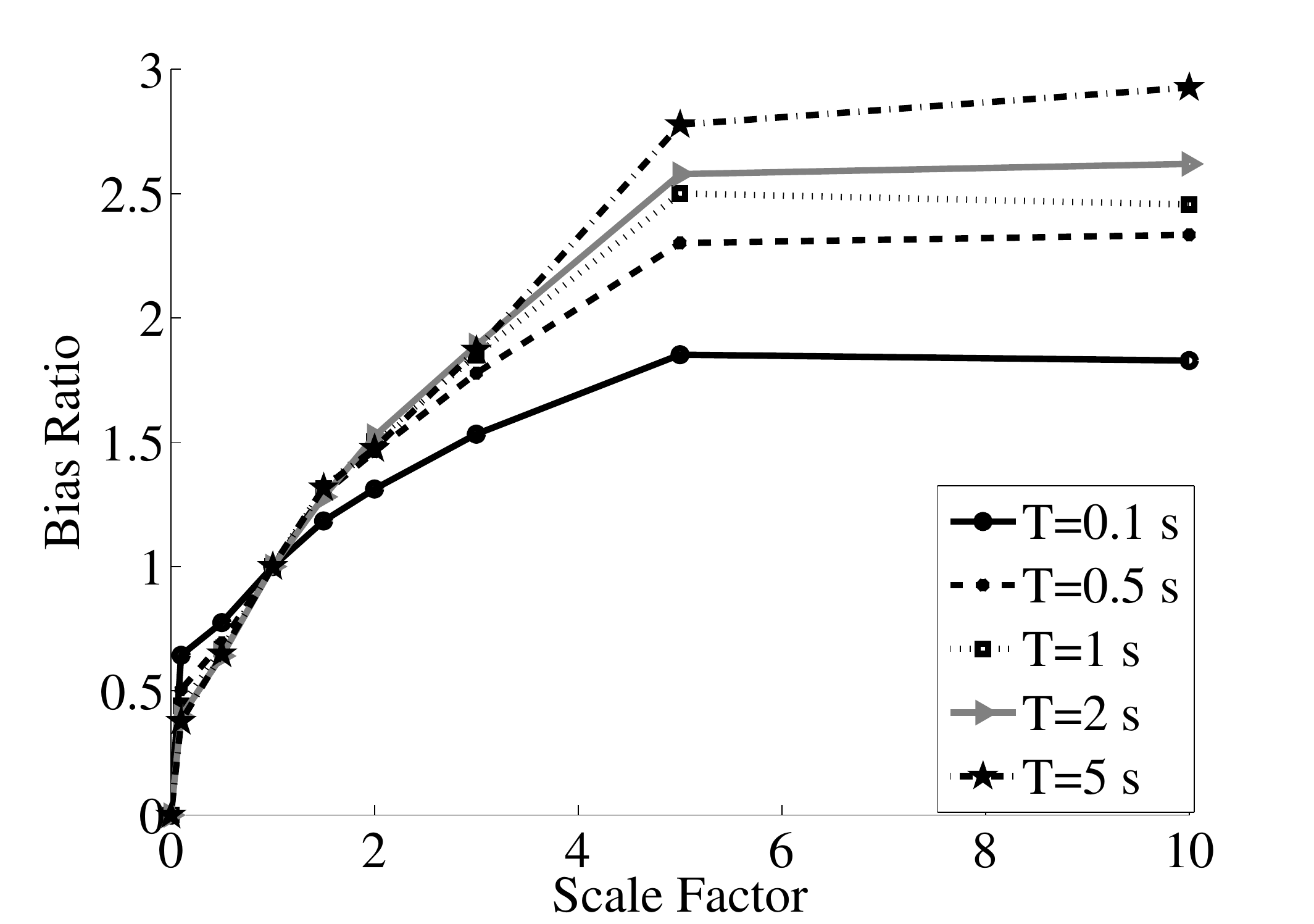}
	}
\subfigure
[Spectral displacement]
	{
		\includegraphics[width=0.45\textwidth]{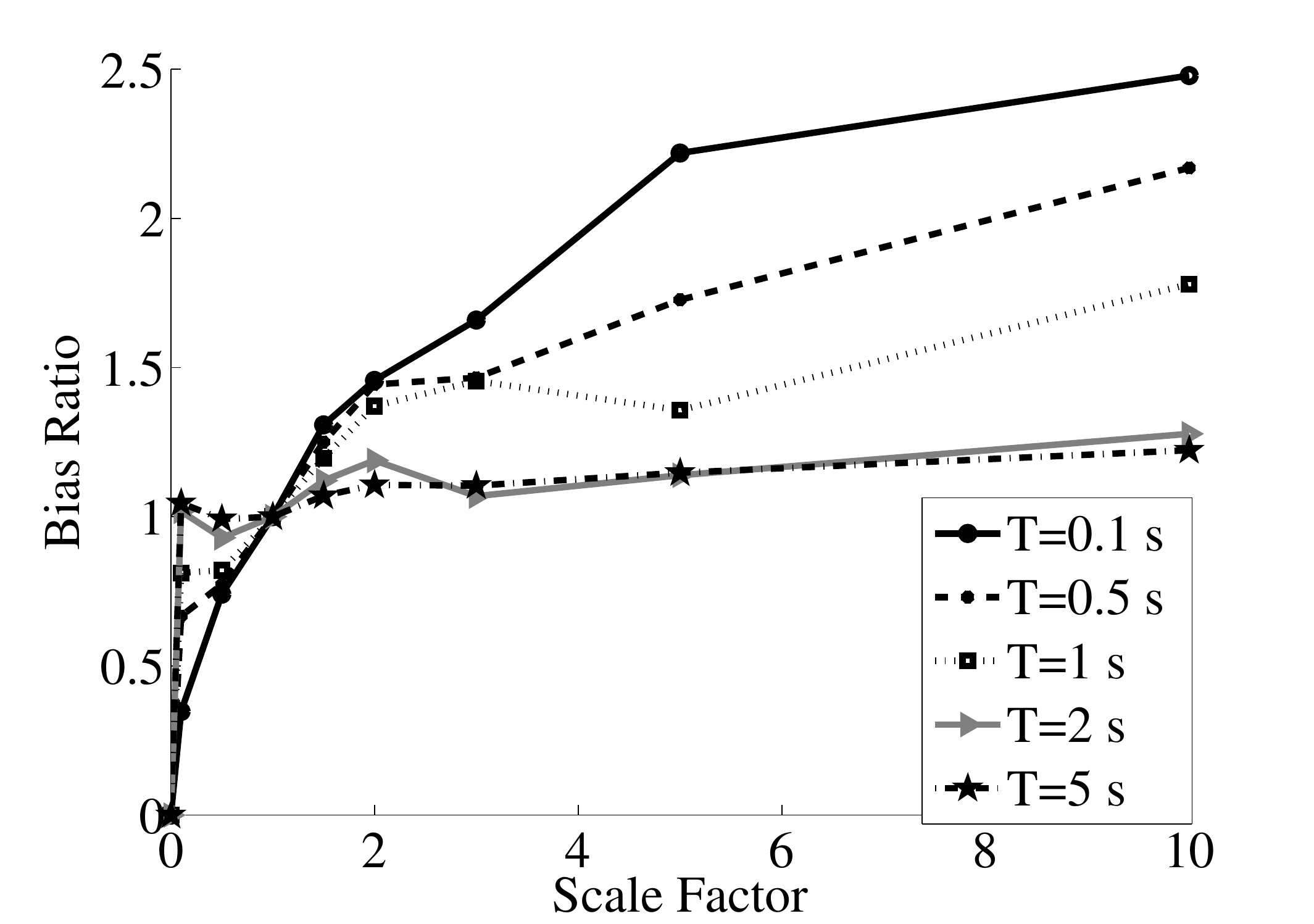}

	}
\caption{Bias ratios induced by the scaling of ground motions for two different intensity measures, namely peak ground acceleration and spectral displacement (after \cite{Mehdizadeh2012}).
}
\label{fig1b}
\end{figure}

The bMCS approach is similar to the incremental
dynamic analysis (IDA) in \cite{Vamvatsikos2002, Vamvatsikos2005, Mander2007} except that when using IDA, one scales \emph{all} the ground motions 
to the intensity level of interest. Therefore in the IDA approach there are signals scaled with very 
large (or very small) scale factors compared to unity which may lead to a gross approximation of the corresponding responses \cite{Mehdizadeh2012, Cimellaro2009}. 
\figref{fig1b} shows the bias ratios induced by the scaling of ground motions for two different intensity measures.
The bias for a certain scale factor is represented by the ratio
of the mean maximal displacement response of a nonlinear SDOF system subject to the scaled motions to the respective response of the system without scaling of the motions.
Note that the bias ratio becomes larger with increasing deviation of the scale factor from unity.
In our approach, \eqrefe{eq13} basically represents the scaling of
the ground motions in the vicinity of the intensity level $IM_o$.
The vicinity is defined by the bin width $2h$ which is chosen so that
the scale factors are close to 1 and typically in the range  [0.8, 1.2] as in the application in Section \ref{sec4}. Accordingly, the bias due to ground motion scaling is negligible.

\subsection{Kernel density estimation}
\label{sec3.3}
The fragility curves defined in \eqrefe{eq7} may be reformulated using the conditional probability 
density function (PDF) $f_{\Delta|IM}$ as follows:
\begin{equation}
 \text{Frag}(a;\,\delta_o)= \Prob{\Delta \geq \delta_o | IM=a}=\int\limits_{ \delta_o}^{+\infty} f_{\Delta}(\delta| IM=a) \, \di \delta
 \label{eq10a}
\end{equation}
By definition this conditional PDF is given as:
\begin{equation}
	f_{\Delta}(\delta|IM=a)=\frac{f_{\Delta,IM}(\delta,a)}{f_{IM}(a)}
	\label{eq10}
\end{equation}
where $f_{\Delta,IM}(\cdot)$ (resp. $f_{IM}(\cdot)$ ) is the joint distribution of the 
vector $(\Delta,\,IM)$ (resp. the marginal distribution of $IM$).
If these quantities were known, the fragility curve in \eqrefe{eq10a} would be obtained by a mere integration.

In this section we propose to \emph{estimate} the joint and marginal PDFs from a sample 
set $\left\{ \prt{IM_i,\Delta_i} \right.$, $\left. i=1 \enu N \right\}$ by \emph{kernel density estimation} (KDE).
For a single random variable $X$ for which a sample set $\acc{x_1 \enu x_N}$ is available, the kernel 
density estimate of the PDF reads \cite{WandJones}:
\begin{equation}
 \hat{f}_X\prt{x}=\dfrac{1}{N h} \sum_{i=1}^N K\prt{\dfrac{x-x_i}{h}}
\end{equation}
where $h$ is the \emph{bandwidth} parameter and $K(\cdot)$ is the \emph{kernel} function which 
integrates to one. Classical kernel functions are the Epanechnikov, uniform, normal and triangular 
functions. The choice of the kernel is known not to affect strongly the quality of the 
estimate \cite{WandJones} provided the sample set is large enough. In case a standard normal PDF 
is adopted for the kernel, \ie $K(x) = \varphi(x) \equiv \expo{-x^2/2}/\sqrt{2\pi}$, the kernel density 
estimate rewrites:
\begin{equation}
 \hat{f}_X\prt{x}=\dfrac{1}{N h} \sum_{i=1}^N \varphi\prt{\dfrac{x-x_i}{h}}
 \label{eq12}
\end{equation}
In contrast, the choice of the bandwidth $h$ is crucial for the kernel density estimate \cite{DuongThesis2004}. An inappropriate 
value of $h$ can lead to an oversmoothed or undersmoothed estimated PDF.
\eqrefe{eq12} is used for estimating the marginal distribution of the $IM$s, namely $\hat{f}_{IM}(a)$ from the set of intensity measures at hand,
namely $\acc{IM_i,\,i=1 \enum N}$:
\begin{equation}
 \hat{f}_{IM}(a)=\dfrac{1}{N h_{IM}} \sum_{i=1}^N \varphi\prt{\dfrac{a-{IM}_i}{h_{IM}}}
\end{equation}

Kernel density estimation may be extended to a random vector $\ve{X} \in \Rr^d$ given 
an i.i.d sample $\acc{\ve{x}_1 \enu \ve{x}_N}$ \cite{WandJones}:
\begin{equation}
 \hat{f}_{\ve{X}}\prt{\ve{x}}=\dfrac{1}{N  \abs{\mat{H}}^{1/2}} \sum_{i=1}^N K\prt{\mat{H}^{-1/2}(\ve{x}-\ve{x}_i)}
\end{equation}
in which $\mat{H}$ is a symmetric positive definite \emph{bandwidth matrix} whose determinant is denoted 
by $\abs{\mat{H}}$. When a multivariate standard normal kernel is 
adopted, the joint distribution estimate becomes:
\begin{equation}
 \hat{f}_{\ve{X}}\prt{\ve{x}}=\dfrac{1}{N \abs{\mat{H}}^{1/2}} \sum_{i=1}^N \dfrac{1}{\prt{2\pi}^{d/2}}\expo{-\frac{1}{2} {(\ve{x}-\ve{x}_i)}\tr \mat{H}^{-1} (\ve{x}-\ve{x}_i)}
 \label{eq14}
\end{equation}
where $(\cdot)\tr$ denotes the transposition.
For multivariate problems (\ie $\ve{X} \in \Rr ^d$), the bandwidth matrix classically
belongs to one of the following classes: spherical, ellipsoidal and full matrix which contains 
respectively 1, $d$ and $d(d+1)/2$ independent unknown parameters. 
This matrix $\mat{H}$ can be computed by means of \eg plug-in and cross-validation
estimators (see \cite{DuongThesis2004} for details). 
In the most general case when the correlations between the random 
variables are not known, the full matrix 
should be used.
In this case, the \textit{smoothed cross-validation estimator} is 
the most reliable among the cross-validation methods \cite{Duong2005}.
Using \eqrefe{eq14} to estimate the joint 
PDF $\hat{f}_{\Delta,IM}(\delta,a)$ from the data set $\acc{(IM_i,\,\Delta_i),\,i=1 \enum N}$ one gets:
\begin{equation}
 \hat{f}_{\Delta,IM}(\delta,a)= \dfrac{1}{2\pi N \abs{\mat{H}}^{1/2}}   \sum\limits_{i=1}^N \expo{-\dfrac{1}{2} {\begin{pmatrix} \delta-\Delta_i\\ a-IM_i \end{pmatrix}}\tr \mat{H}^{-1}  \begin{pmatrix} \delta-\Delta_i\\ a-IM_i \end{pmatrix} }
\label{eq15}
 \end{equation}

The conditional PDF $f_{\Delta}(\delta|IM=a)$ is eventually estimated by pluging the estimations of the numerator and denominator in \eqrefe{eq10}.
The proposed estimator of the fragility curve eventually reads:
\begin{equation}
  \widehat{\text{Frag}}(a;\,\delta_o)
	=  \dfrac{h_{IM}}{2\pi \abs{\mat{H}}^{1/2}} \dfrac  {  \int\limits_{ \delta_o}^{+\infty}  \sum\limits_{i=1}^N \expo{-\dfrac{1}{2} {\begin{pmatrix} \delta-\Delta_i\\ a-IM_i \end{pmatrix}}\tr \mat{H}^{-1}  \begin{pmatrix} \delta-\Delta_i\\ a-IM_i \end{pmatrix} } \text{d}\delta}
	{ \sum\limits_{i=1}^N \varphi\prt{\dfrac{a-IM_i}{h_{IM}}}}
 \label{eq17}
\end{equation}

The choice of the bandwidth parameter $h$ and the bandwidth matrix $\mat{H}$ plays a crucial role in 
the estimation of fragility curves, as seen in \eqrefe{eq17}.
In the above formulation, the same bandwidth is considered for the whole range of the $IM$ values. 
However, there are typically few observations corresponding to the upper tail of the distribution of $IM$.
This is due to the fact that the annual frequency of seismic motions with $IM$ values in the respective range (\eg $PGA$ exceeding $1g$) is low (see \eg \cite{Frankel2000}). 
This is also the case when synthetic ground motions are used, since these 
are generated consistently with the statistical features of recorded motions. 
Preliminary investigations have shown that by applying the KDE method
on the data in the original scale, the fragility curves for the higher demand thresholds tend to be unstable in their upper tails \cite{SudretMaiCFM2013}.
To reduce effects of the scarcity of observations at large $IM$ values, we propose the use of KDE in the logarithmic scale, as described next.

Let us consider two random variables $X$, $Y$ with positive supports and their logarithmic transformations  $U= \log X$ and
$V= \log Y$.
The probability contained in a differential area must be invariant under the change of variables. In the
one-dimensional case, this leads to:
\begin{equation}
 f_X(x) = \abs{\dfrac{\di u}{\di x}} f_U(U)  = \abs{\dfrac{1}{x}} f_U(u) = \dfrac{1}{x} f_U(u)
 \label{eq22}
\end{equation}
In the two-dimensional case, one obtains:
\begin{equation}
 f_{X,Y}(x,y) = \dfrac{1}{x\,y} f_{U,V}(u,v)
 \label{eq25}
\end{equation}
Using Equations (\ref{eq22}) and (\ref{eq25}), one has:
\begin{equation}
  \int\limits_{y_0}^{+\infty} f_Y(y | X=x) \, \di y 
  =
  \int\limits_{y_0}^{+\infty} \dfrac{f_{X,Y}(x,y)}{f_X(x)} \, \di y 
  = 
  \int\limits_{\log y_0}^{+\infty} \dfrac{\dfrac{1}{x\,y} f_{U,V}(u,v)}{\dfrac{1}{x} f_U(u)} \, y \, \di v
  =
  \int\limits_{\log y_0}^{+\infty} f_V (v | U=u) \, \di v
 \label{eq26}
\end{equation}
Accordingly, by considering $X =IM$ and $Y=\Delta$, the fragility curves defined in \eqrefe{eq10a} can be estimated in terms of $U=\log IM$ and $V=\log \Delta$ as:
\begin{equation}
 \widehat{\text{Frag}}(a;\,\delta_o)=\int\limits_{ \delta_o}^{+\infty} \hat{f}_{\Delta}(\delta| IM=a) \, \di \delta
 = \int\limits_{\log \delta_o}^{+\infty} \hat{f}_V (v | U= \log a) \, \di v
 \label{eq27}
\end{equation}

Note that use of a constant bandwidth in the logarithmic scale 
is equivalent to the use of a varying bandwidth in the original scale, 
with larger bandwidths corresponding to larger values of $IM$. 
The resulting fragility curves are smoother than 
those obtained by applying KDE with the data in the original scale.

\subsection{Epistemic uncertainty of fragility curves}
\label{sec3.4}

It is of utmost importance in fragility analysis to investigate the variability in the estimated curves arising due to epistemic uncertainty.
This is because any fragility curve is always computed based on a limited amount of data, \ie a limited number of ground motions and related structural analyses. Large epistemic uncertainties may affect significantly
the total variability of the seismic risk assessment outcomes. Consequently, characterizing and propagating epistemic uncertainties in seismic loss estimation has attracted attention from several researchers \cite{Baker2008a,Bradley2010,Liel2009}.

The theoretical approach to determine the variability of an estimator relies on repeating the estimation with an ensemble of different random samples. However, this approach is not feasible in earthquake engineering because of the high computational cost. In this context, the \textit{bootstrap resampling} technique \cite{Efron1979} is deemed appropriate \cite{Baker2008a}. Given a set of observations $\ve{X}=\prt{X_1 \enum X_n}$ from an unspecified probability distribution $F$, the bootstrap method allows estimation of the statistics of a random variable that depends on both $\ve{X}$ and $F$ in terms of the observed data $\ve{X}$ and their empirical distribution.

To estimate statistics of the fragility curves with the bootstrap method, we first draw $M$ independent random samples \emph{with replacement} from the original data set $\acc{\prt{IM_i,\Delta_i}, i=1 \enu N}$. These represent the so-called bootstrap samples. Each bootstrap sample has the same size $N$ as the original sample, but the observations are different: in a particular sample, some of the original observations may appear multiple times while others may be missing. Next, we compute the fragility curves for each bootstrap sample using the approaches in Sections \ref{sec3.1}, \ref{sec3.2} and \ref{sec3.3}. Finally, we perform statistical analysis of the so-obtained $M$ bootstrap curves. 
In the subsequent example illustration, the above procedure is employed to evaluate the median and 95\% confidence intervals of the estimated fragility curves and also, to assess the variability in the $IM$ value corresponding to a $50 \%$ probability of failure.

In the following, the proposed non-parametric approaches, namely bMCS and KDE, as well as the classical lognormal approach for estimation of fragility curves are demonstrated in an example application on a frame structure. The uncertainty in the estimation of the bMCS- and KDE-based curves is also investigated.

\section{Illustration: steel frame structure}
\label{sec4}
\subsection{Problem statement: 3-storey 3-span steel frame}
\begin{figure}[!ht]
\centering
\subfigure
	{
		\includegraphics[width=0.5\textwidth]{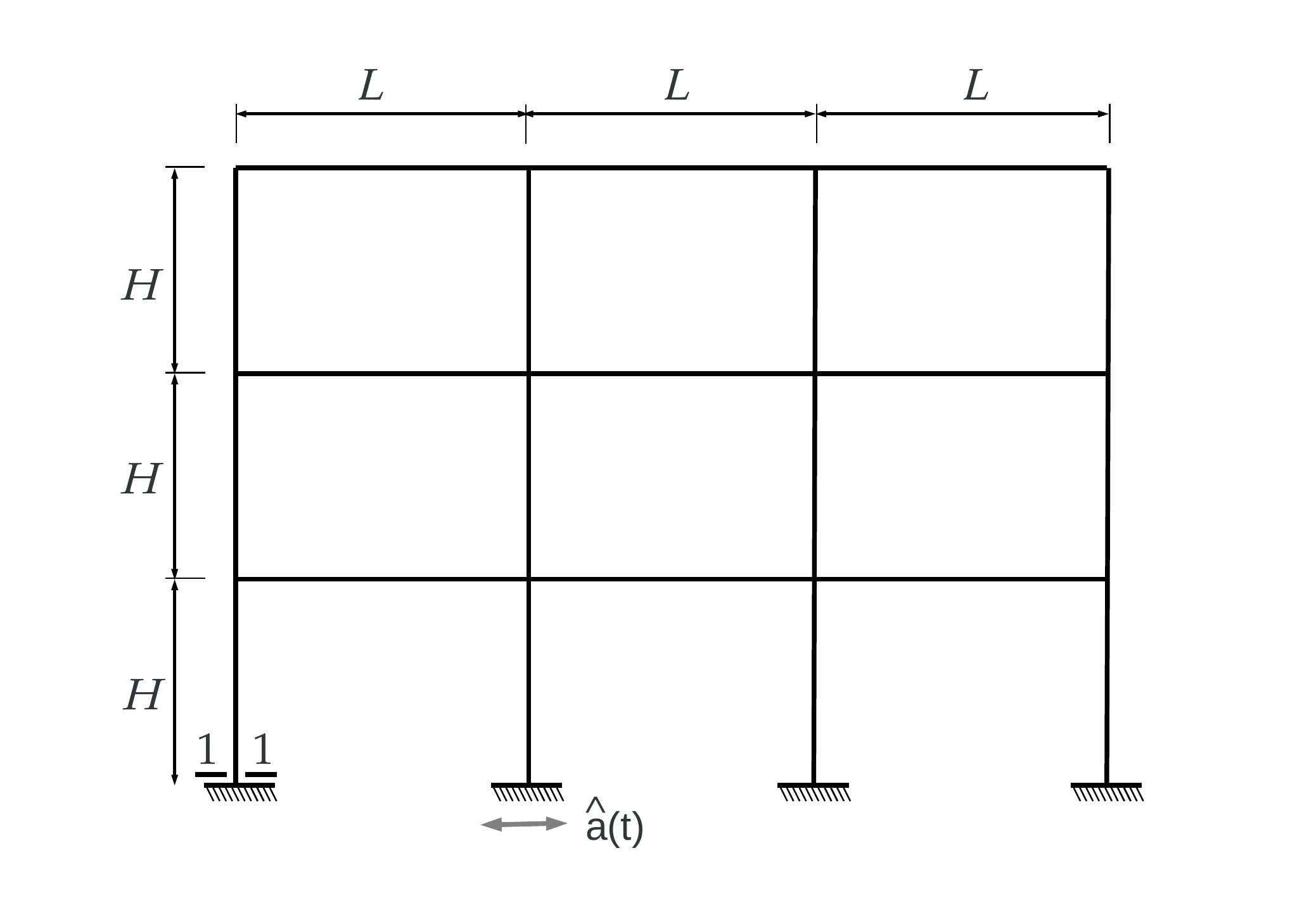}
	}
\subfigure
	{
		\includegraphics[width=0.4\textwidth]{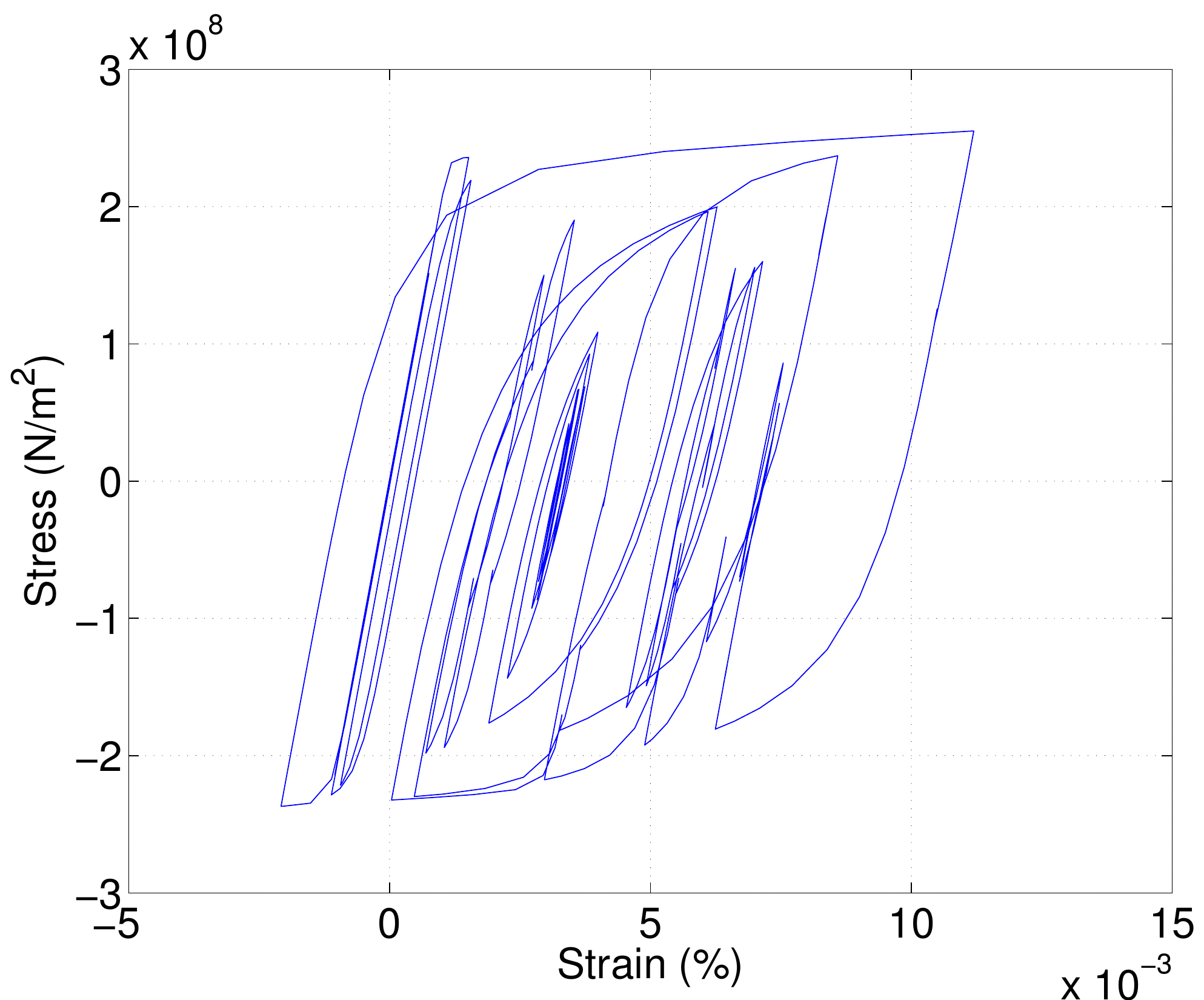}
	}
\caption{(Left) Steel frame structure. (Right) Hysteretic behavior of steel material at section 1-1 when the frame is subject to a particular ground motion.}
\label{structure}
\end{figure}

\begin{figure}[!ht]
\centering
\includegraphics[width=0.25\textwidth]{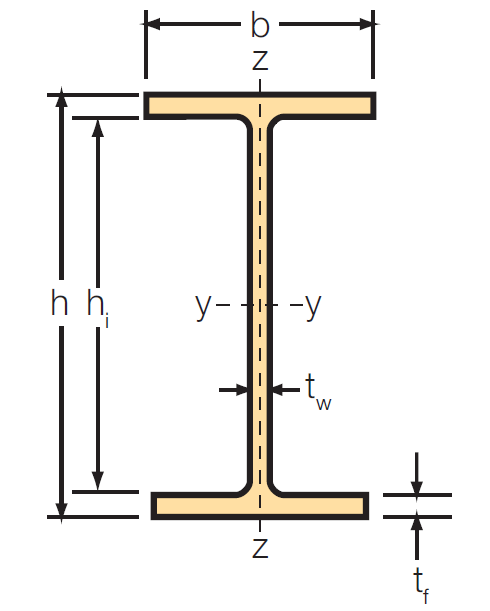}
\caption{Cross-sections of beams ($h$=306~mm, $b$=147~mm, $t_w$=8,5~mm, $t_f$=13,7~mm) and columns ($h=336$~mm, $b$=158~mm, $t_w$=9,2~mm, $t_f$=14,5~mm).}
\label{sections}
\end{figure}

We evaluate the fragility curves of a 3-storey 3-span steel frame structure
with the following dimensions: storey-height $H=3$~m, span-length $L=5$~m (see \figref{structure}). The 
steel material has nonlinear isotropic hardening behavior following the uniaxial
Giuffre-Menegotto-Pinto steel model as implemented in the finite element software 
OpenSees \cite{OpenSees}. \cite{Ellingwood2009} have shown that uncertainty in the properties of the steel material has a negligible effect on seismic fragility curves. Therefore, mean material properties are employed in the subsequent fragility analysis. In particular, the properties considered are $E_0= 210,000$~MPa for the Young's modulus, $f_y=264$~MPa for the yield strength \cite{EC3, jcss} and $b=0.01$ for the strain 
hardening ratio (ratio of the post-yield tangent to the initial value). 
\figref{structure} depicts the hysteretic behavior of the steel material at 
a section of the frame for an example ground motion. 
The structural components are modelled with nonlinear elements with distributed plasticity along their lengths.
The use of fiber sections allows modelling the plasticity over the element cross-sections \cite{Deierlein2010}. 
The vertical load consists of dead-load (from the frame elements 
as well as the supported floors), and live load in accordance with Eurocode~1~\cite{EC1} which result in a
total distributed load on the beams $q=20$~kN/m. The preliminary design is performed using the vertical loads to
provide the necessary sections of columns and beams.
The standard European I beams with designation IPE 300 R and IPE 330 R are chosen respectively for beams and columns.
The dimensions of the sections are depicted in \figref{sections}.
The first two natural periods (resp. frequencies) of the building obtained by modal analysis of the corresponding linear model
are $T_1=0.42$~s and $T_2=0.13$~s (resp. $f_1=2.38$~Hz and $f_2=8.33$~Hz).

\begin{figure}[!ht]
\centering
\subfigure
	{
		\includegraphics[width=0.45\textwidth]{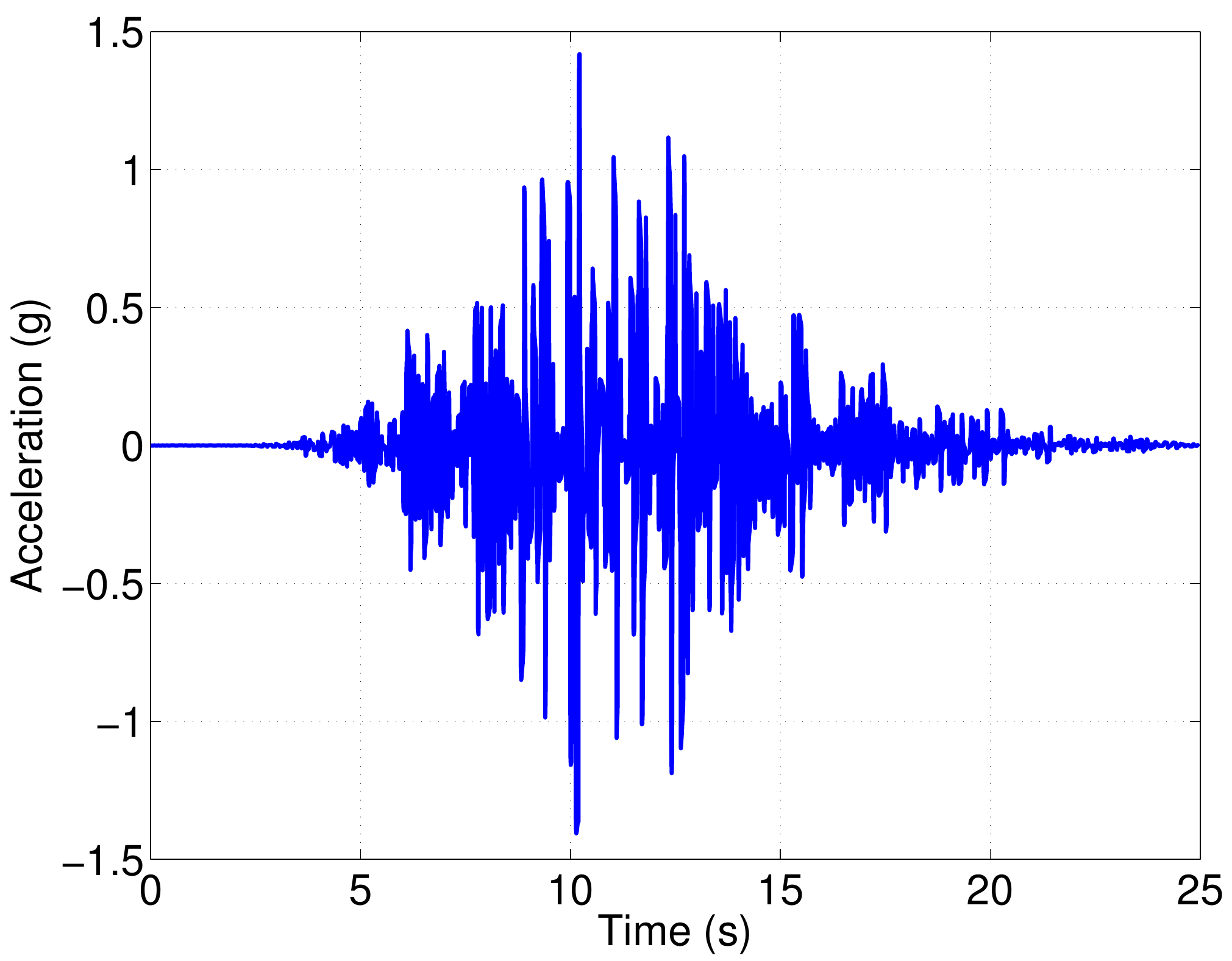}
	}
\subfigure
	{
		\includegraphics[width=0.45\textwidth]{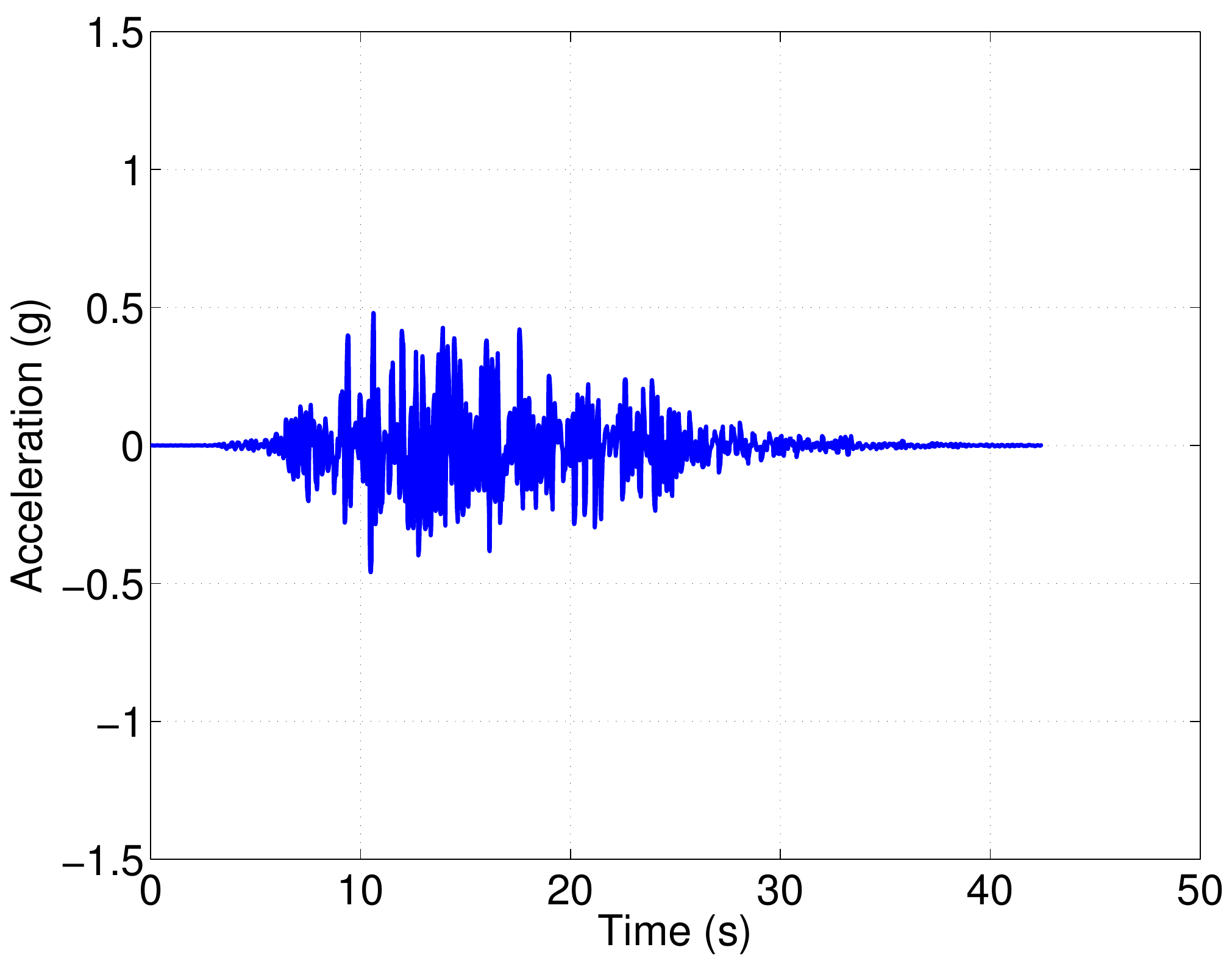}
	}
\subfigure
	{
		\includegraphics[width=0.45\textwidth]{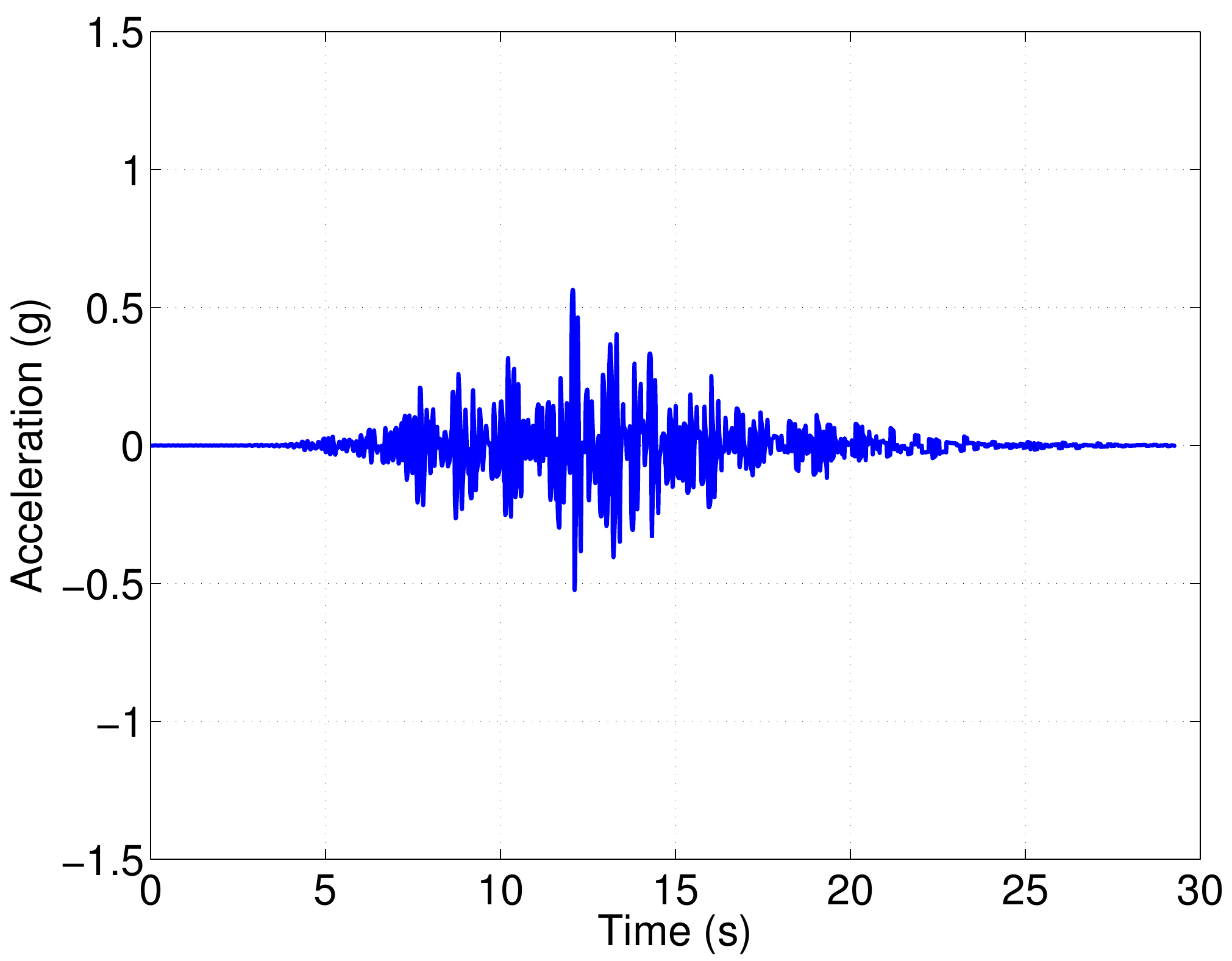}
	}
\subfigure
	{
		\includegraphics[width=0.45\textwidth]{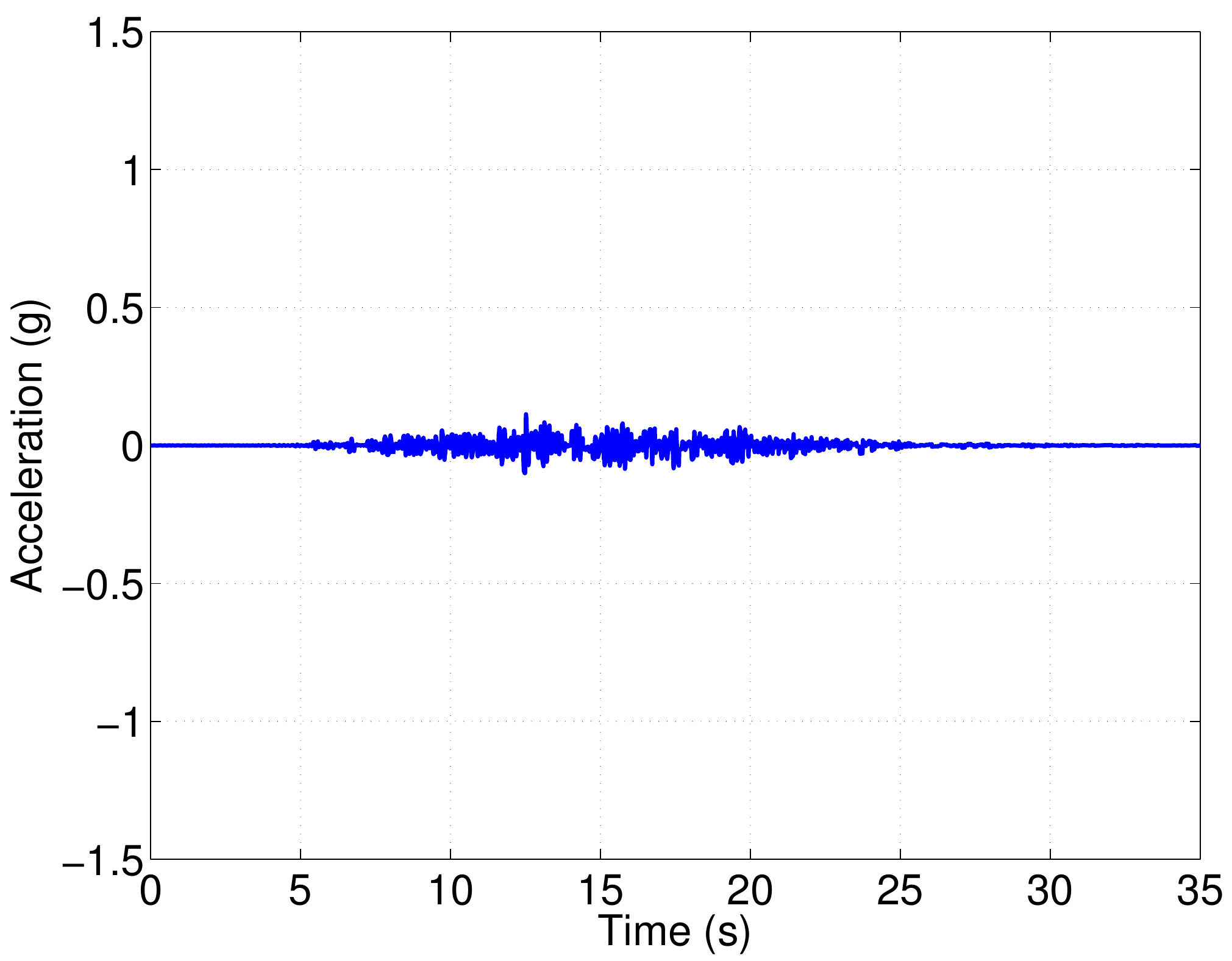}
	}
		\caption{Examples of generated synthetic ground motions.}
		\label{fig4}
\end{figure}

The structure is subject to ground motions represented by the acceleration time histories
at the ground level. Each ground motion is modelled by 
6 randomized parameters $\prt{\alpha_1,\alpha_2,\alpha_3,\omega_{mid}, \omega', \zeta_f }$ 
directly related to the parameters in Table~\ref{tab:1} and a Gaussian input 
vector $\ve{U}$ (\eqrefe{eq:at2}). The reader is referred to \cite{Rezaeian2010} for viewing the correlation between these parameters.
Note that in order to obtain a sufficiently wide range of $IM$, we adapted 
the statistics of $I_a$, $D_{5-95}$ and $t_{mid}$ presented in \cite{Rezaeian2010}. The 
duration of each synthetic earthquake is computed based on the corresponding set 
of parameters $\prt{\alpha_1,\alpha_2,\alpha_3}$ and is used to
determine the size of the Gaussian vector $\ve{U}$. A set of four different 
synthetic ground motions is shown in \figref{fig4} for the sake of illustration.
The finite element code OpenSees \cite{OpenSees} is used
to carry out transient dynamic analyses of the frame for a total of $N=10^4$ synthetic ground motions. 

Numerous $IM$s can be used to represent the earthquake severity, see \eg \cite{Mackie2004}.
Peak ground acceleration ($PGA$) is a convenient measure that 
is straightforward to obtain for a given ground motion and
has been traditionally used in attenuation relationships and design codes. 
However, structural responses may exhibit large dispersions with respect to $PGA$, 
since they are also highly dependent on other features of earthquake motions, 
e.g. the frequency content and the duration of the strong motion phase. 
Structure-specific $IM$s, such as the spectral acceleration $Sa$, 
tend to be better correlated with structural responses \cite{Mackie2004,Padgett2008a}. 
In the following, we compute fragility curves considering both $PGA$ and $Sa$ as $IM$s. 
In the present example, $Sa$ represents $Sa(T_1)$ \ie the spectral acceleration for 
a SDOF system with period equal to the fundamental period $T_1$ of the frame and viscous damping factor equal 5\%.

The engineering demand parameter commonly considered in fragility analysis of steel buildings is the inter-storey drift ratio (see \eg \cite{Ellingwood2009,Cornell2002,Lagaros2007}). Accordingly, we herein develop fragility curves for three different thresholds of the maximum inter-storey drift ratio over the building. To gain insight into structural performance, we consider the thresholds 0.7\%, 1.5\% and 2.5\%, which are associated with different damage states in seismic codes. In particular, the thresholds 0.7\% and 2.5\% are recommended by \cite{FEMA2000} to respectively characterize light damage and moderate damage for steel frames. The drift threshold 1.5\% corresponds to the damage limitation requirement for buildings with ductile non-structural elements according to \cite{EC8eng}. These descriptions only serve as rough damage indicators herein, since in the PBEE framework the relationship between drift limit and damage is probabilistic.

\begin{table}[!ht]
\caption{Statistics of parameters used to generate the synthetic ground motions (adapted from~\cite{Rezaeian2010}).}
\centering
\begin{tabular}{m{0.15\textwidth} m{0.15\textwidth} m{0.15\textwidth} m{0.15\textwidth} m{0.15\textwidth}}
\toprule
Parameter & Distribution & Support & $\mu_X$ & $\sigma_X$ \\
\midrule
$I_a$ & Lognormal & (0, +$\infty$) & 5.613 & 10.486 \\
$D_{5-95}$ & Beta      & [5, 20]  & 10 & 2 \\
$t_{mid}$ & Beta & [5, 15] & 12 & 2 \\
$\omega_{mid}$ / {2$\pi$} & Gamma & (0, +$\infty$) & 5.930 & 3.180 \\
$\omega'$ / {2$\pi$} & Two-sided exponential&[-2, 0.5] &-0.089 & 0.185 \\
$\zeta_f$ &Beta & [0.02, 1] & 0.210 & 0.150\\
\bottomrule
\end{tabular}
\label{tab:1}
\end{table}

\subsection{Results}
\begin{figure}[!ht]
 \centerline{
\includegraphics[width=0.6\textwidth]{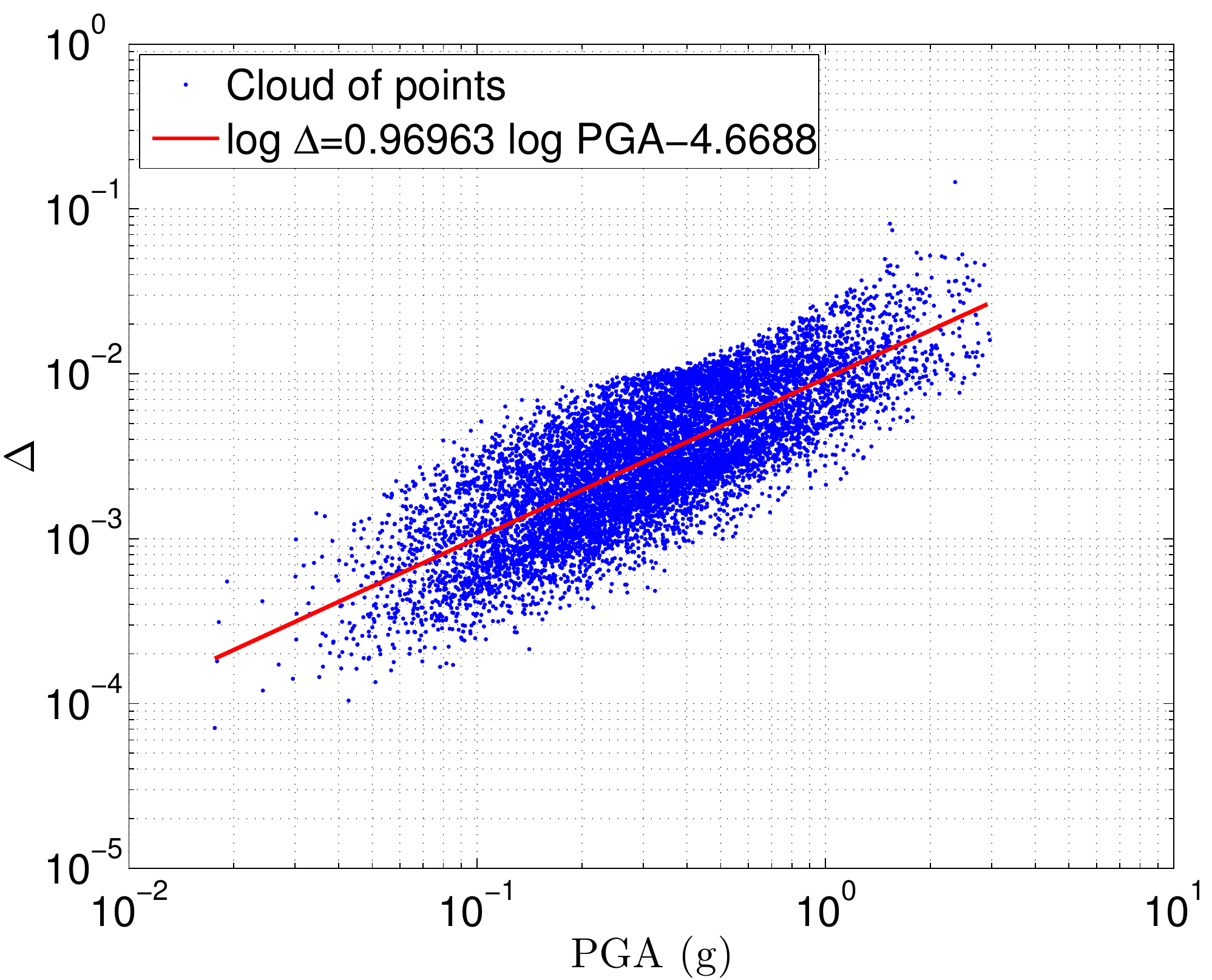}
}
\caption{Paired data $\acc{\prt{PGA_i,\Delta_i}, i=1 \enu 10^4}$ and linear regression in the logarithmic scale.}
\label{regression}
\end{figure}

\begin{figure}[ht!]
 \centerline{
\includegraphics[width=0.6\textwidth]{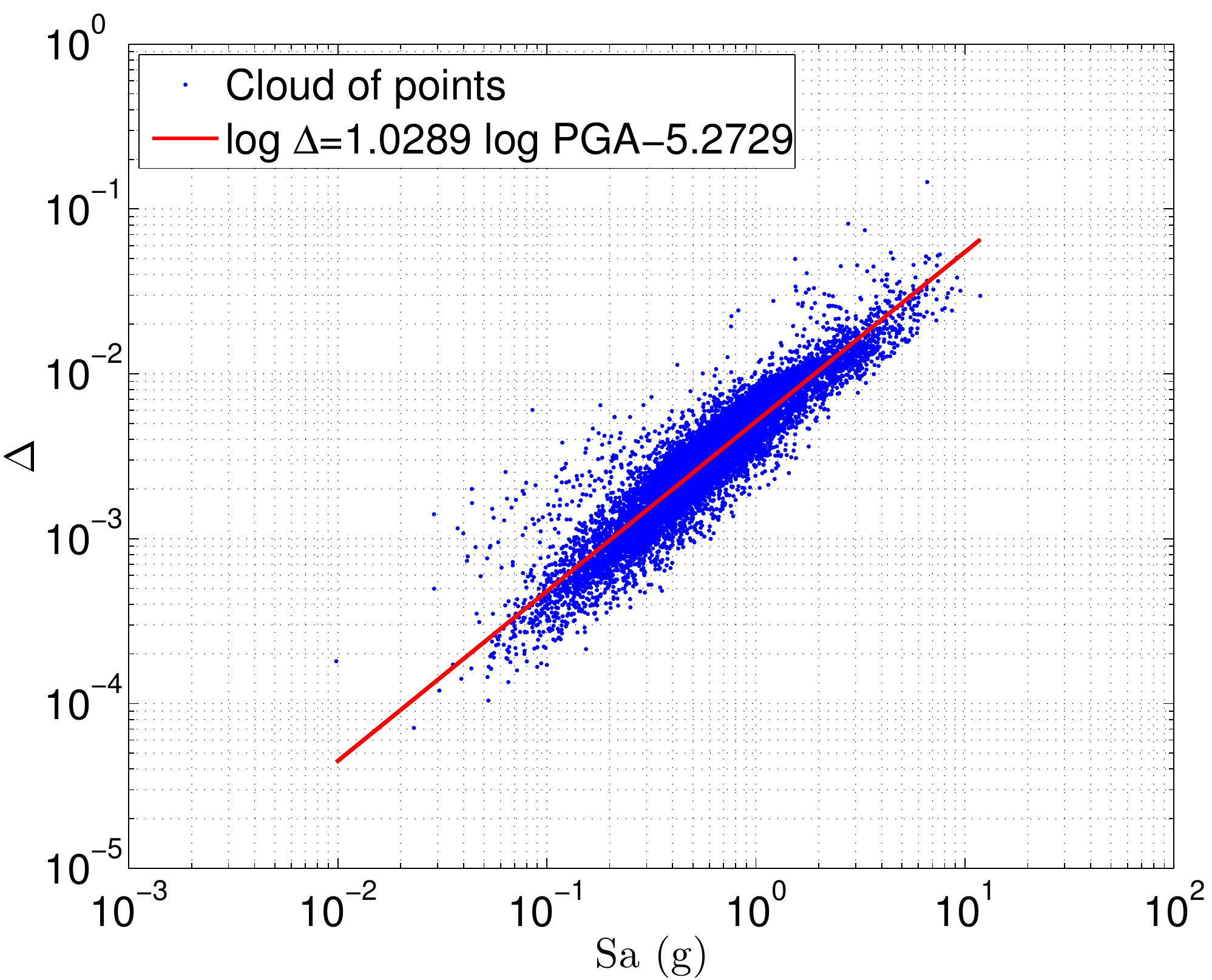}
}
 \caption{Paired data $\acc{\prt{Sa_i,\Delta_i}, i=1 \enu 10^4}$ and linear regression in the logarithmic scale.}
 \label{fig7}
\end{figure}

As described in Section \ref{sec3}, the lognormal approach consists in assuming that the fragility curves have the shape of a lognormal CDF and then estimating the parameters of the distribution. Using the MLE approach, the observed failures for each drift threshold are modeled as outcomes of a Bernoulli experiment and the parameters of the fragility curves are determined by maximizing the respective likelihood function. Using the linear regression technique, the parameters of the lognormal curves are indirectly derived by fitting a linear model to the paired data $(\log IM,\,\log \Delta)$. \figref{regression} (resp. \figref{fig7}) depicts the paired data $(\log PGA,\,\log \Delta)$ (resp. $(\log Sa,\,\log \Delta)$) and the fitted PSDM using linear regression. The coefficient of determination $R^2$ of the linear regression model considering $PGA$ (resp. $Sa$) as an intensity measure is 0.607 (resp. 0.866). We observe that using $Sa$ as an intensity measure leads to smaller dispersion \ie smaller $\zeta$ in \eqrefe{eq8} as compared to using $PGA$. This is expected 
since $Sa$ is a structure-specific $IM$. In the bMCS method, the bandwidth $h$ is set equal to $0.2\,IM_o$. The resulting scale factors vary in the range $[0.8,\,1.2]$ yielding a bias ratio of approximately 1 (see \figref{fig1b}). Finally, the KDE approach requires estimation of the bandwidth parameter and the bandwidth matrix. Using the smoothed cross-validation estimation implemented in \texttt{R} \cite{Duongks2007}, these are determined as $h= 0.1295$ and $\ve{H}= \begin{bmatrix}
0.0306 & 0.0246 \\
0.0246 & 0.0283
\end{bmatrix}$ (resp. $h=0.1441$ and 
$\ve{H}=\begin{bmatrix}
0.0229 & 0.0224 \\
0.0224 & 0.0248
\end{bmatrix}$) when $PGA$ (resp. $Sa$) is considered as $IM$. 

For the two considered $IM$s and the three drift limits of interest, \tabref{tab:2} lists the medians and log-standard deviations of the lognormal curves obtained with both the MLE and linear regression (LR) approaches. The median determines the position where the curve attains the value 0.5, whereas the log-standard deviation is a measure of the steepness of the curve. The medians of the KDE-based curves have been computed and are also listed in \tabref{tab:2} for comparison. Larger deviations of the lognormal medians from the reference KDE-based values are observed at the larger drift thresholds for both parametric approaches. Note that the MLE approach yields a distinct log-standard deviation for each drift threshold, whereas a single log-standard deviation for all drift thresholds is obtained with the linear regression approach. The standard deviations obtained with MLE may be smaller or larger than those obtained with linear regression depending on the considered $IM$ and threshold.

\begin{table}[!ht]
\caption{Steel frame structure - Parameters of the obtained fragility curves. }
\centering
\begin{tabular}{cccccc}
\toprule
  & & \multicolumn{2}{c}{PGA}  & \multicolumn{2}{c}{Sa}    \\
\cline{3-6}
$\delta_o$ & Approach & Median & Log-std & Median & Log-std  \\
\midrule  
\multirow{3}{*}{0.7\%}  
  & MLE & $0.6844\,g$ &  0.6306 & $1.2327\,g$ &  0.2915  \\
& LR & $0.7392\,g$ & 0.6260 & $1.3529\,g$ & 0.3433 \\
  & KDE & $0.7\,g$ &  & $1.2\,g$ &  \\
\cline{2-6}
\multirow{3}{*}{1.5\%}  
 & MLE & $1.7437\, g$ &  0.5194 & $3.4184\,g$ &  0.4327   \\
& LR & $1.6222\,g$ & 0.6260 & $2.8377\,g$ & 0.3433  \\
 & KDE & $1.8\,g$ & & $3.3\,g$ &     \\
 \cline{2-6}
 \multirow{3}{*}{2.5\%}  
  & MLE & $2.4958\,g$ &  0.4631 & $6.3357\,g$ &  0.3769  \\
 & LR & $2.7472\,g$ & 0.6260 & $4.6621\,g$ & 0.3433 \\
 & KDE & $3.0\,g$ & & $5.6\,g$ &     \\

\bottomrule
\end{tabular}
\label{tab:2}
\end{table}

\begin{figure}[ht!]
 \centerline{
\includegraphics[width=0.6\textwidth]{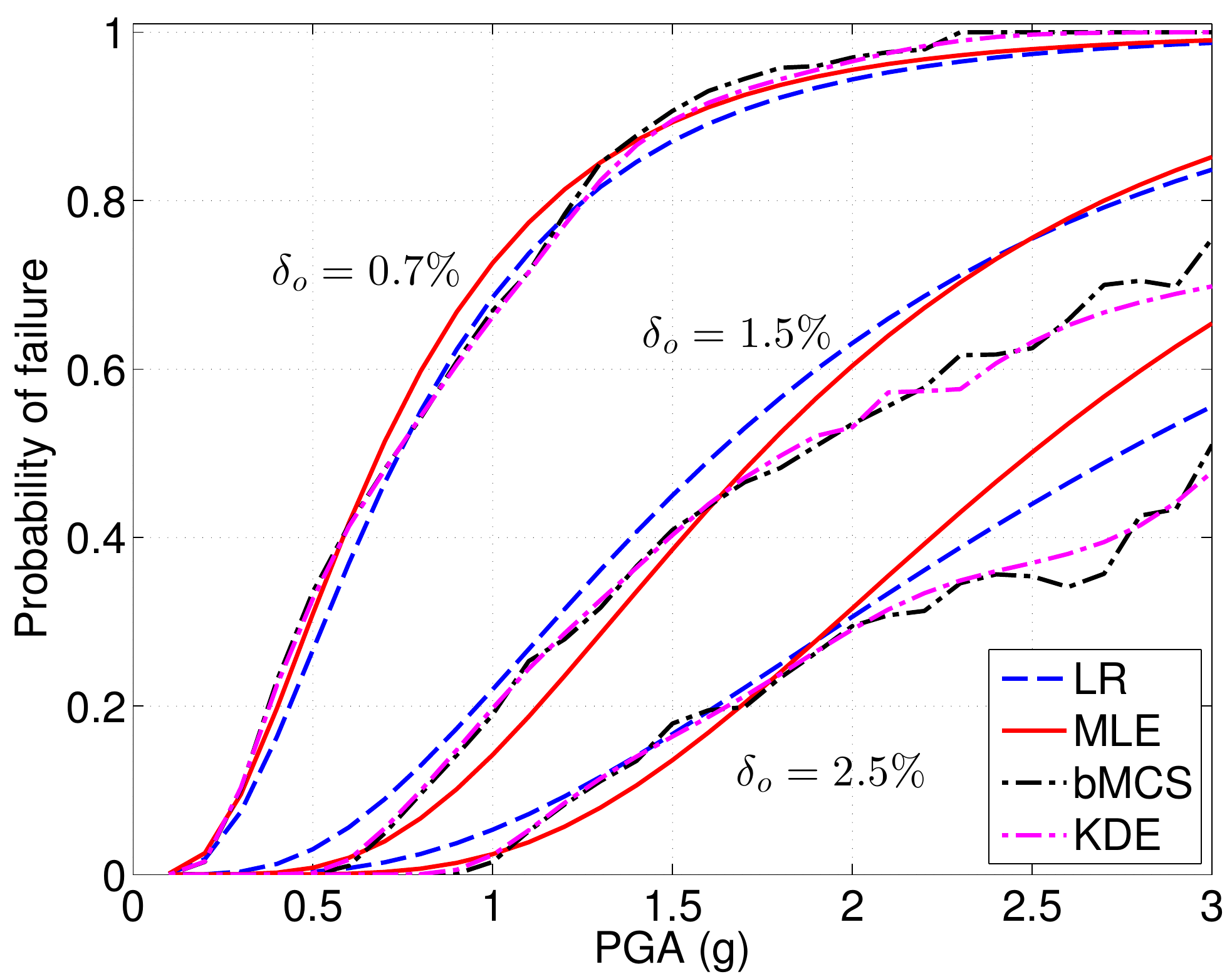}
}
 \caption{Fragility curves by parametric and non-parametric approaches using $PGA$ as intensity measure. (LR: linear regression; MLE: maximum likelihood estimation; MCS: binned Monte Carlo simulation; KDE: kernel density estimation)}
 \label{fig5}
\end{figure}

For the three considered drift thresholds, \figref{fig5} shows the lognormal (with two estimation approaches) and non-parametric (both bMCS- and KDE-based) fragility curves when $PGA$ is used as $IM$.
One first observes a remarkable consistency between the curves obtained with the two proposed non-parametric 
approaches for all limit states although the algorithms 
behind the methods are totally different. This validates the accuracy of the proposed approaches. 
For high values of 
$PGA$ ($>2\,g$), some noise is observed on the bMCS curves due to the 
scarcity of observations in the corresponding intervals. This noise 
may be reduced by using additional observations in this range of $PGA$. 
Note, however, that the set of synthetic motions used in our analysis 
has statistical features characterizing the occurrence of actual earthquake motions
(ground motions with large $IM$s have relatively low probabilities of occurrence).
For the lower threshold ($\delta_o=0.7\%$), the parametric curves are in fairly good agreement with the non-parametric curves. For the two higher thresholds, significant discrepancy is observed between the lognormal curves and the non-parametric ones. In particular, in the range of high $PGA$ values, both lognormal curves overestimate the failure probabilities. It is noted that for $\delta_o=2.5\%$, the median $PGA$ (leading to 50\% probability of exceedance) is respectively underestimated by 9\% and 17\% with the MLE and the linear regression aproach, respectively.

\begin{figure}[ht!]
 \centerline{
\includegraphics[width=0.6\textwidth]{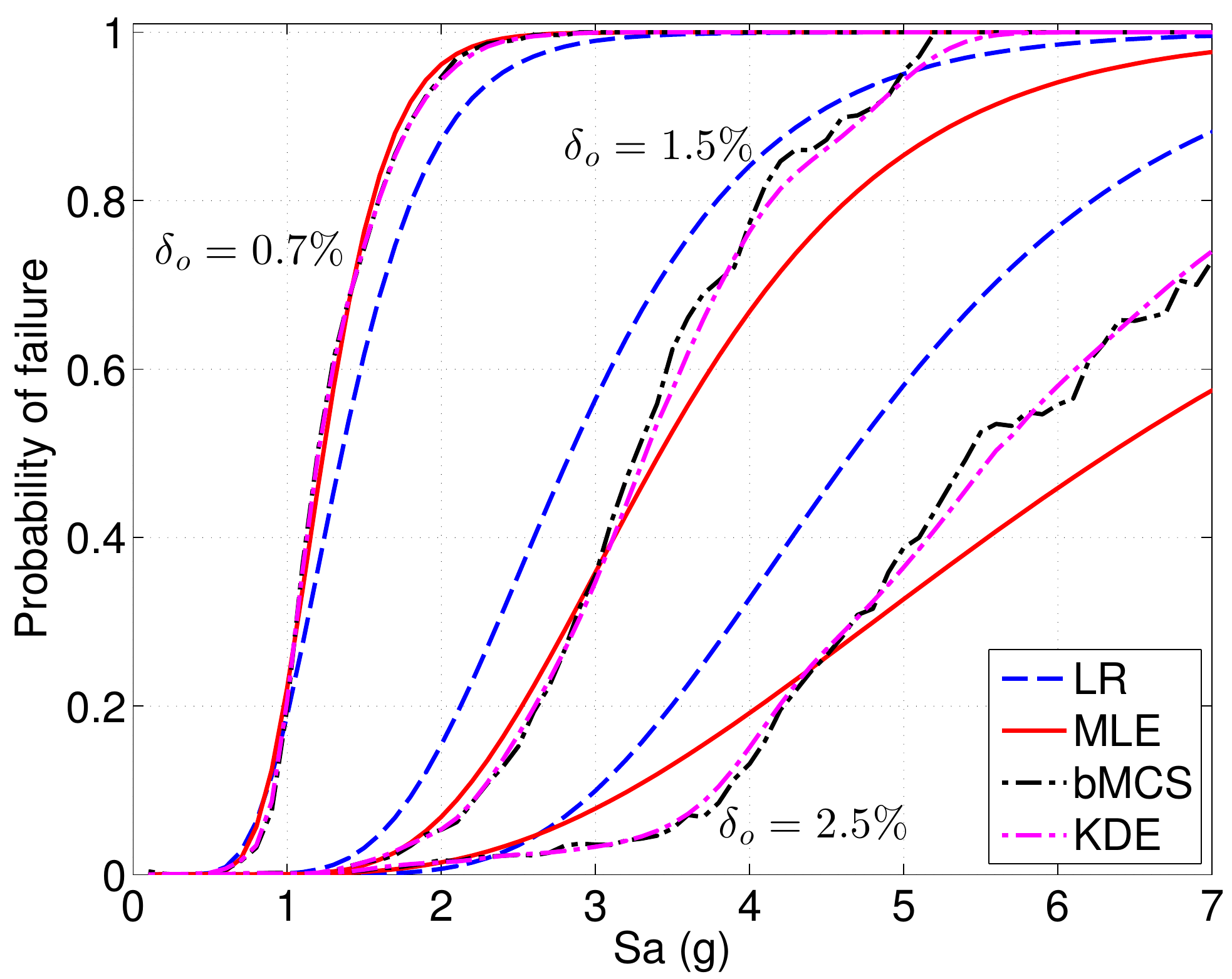}
}
 \caption{Fragility curves by parametric and non-parametric approaches using $Sa$ as intensity measure. (LR: linear regression; MLE: maximum likelihood estimation; MCS: binned Monte Carlo simulation; KDE: kernel density estimation)}
 \label{fig8}
\end{figure}

\figref{fig8} shows the resulting fragility curves when $Sa$ is used as $IM$.
The non-parametric curves by bMCS and KDE remain consistent independently of the limit state. 
For the two larger thresholds, the lognormal curves exhibit large variations from the non-parametric curves as well as between each other. In particular, the curves obtained with MLE underestimate moderate and large failure probabilities whereas the curves obtained with linear regression tend to overestimate failure probabilities. 
For $\delta_o=2.5\%$, the median $PGA$ is overestimated by 13\% with the MLE approach and is underestimated by 17\% with the linear regression approach.
Note that by comparing the absolute discrepancies, linear regression estimation provides less accurate curves for $Sa$ than for $PGA$, although the $R^2$ coefficient of the linear fit is higher for the former. This is due to the fact that the assumption of \textit{normally distributed errors with constant variance}, which is inherent in \eqrefe{eq8}, is not valid for $Sa$, as one observes in \figref{fig7}. Under the assumption of a \textit{unique linear model}, the median is overestimated at the lower drift limit and underestimated at the two larger drift limits.

The previous analysis shows that the two non-parametric approaches, namely bMCS and KDE, yield consistent fragility curves for both $IM$s and for all considered limit states. Using the non-parametric fragility curves as reference, the accuracy of the lognormal curves is found to depend strongly on the method for estimating the parameters of the underlying CDF, the considered $IM$ and the drift threshold of interest. 
Note that different $IM$s have been recommended in literature \cite{Padgett2008a, Jankovic2004} for structures of different type, size and material. Accordingly, the accuracy of the lognormal fragility curves may depend on those factors as well. Possible dependency of the accuracy of the lognormal curves on the considered response quantity also needs to be investigated. 

\subsection{Estimation of epistemic uncertainty by bootstrap resampling}

In the following, we use the bootstrap resampling technique (see Section \ref{sec3.4}) to investigate the epistemic uncertainty in the fragility curves estimated with the proposed non-parametric approaches.

\begin{figure}[ht!]
\centering
\subfigure
[Binned Monte Carlo simulation]
	{
		\includegraphics[width=0.45\textwidth]{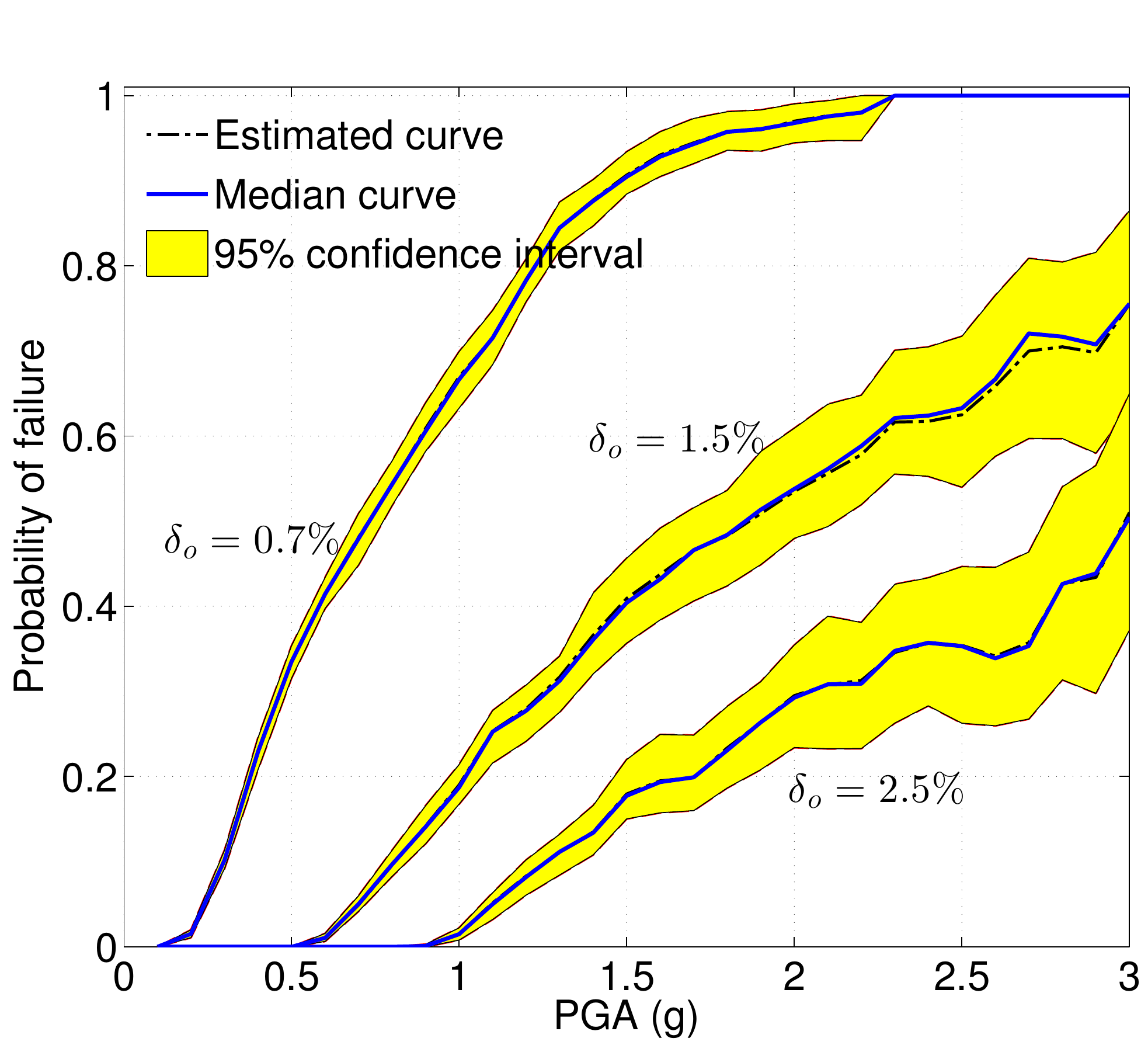}
	}
\subfigure
[Kernel density estimation]
	{
		\includegraphics[width=0.45\textwidth]{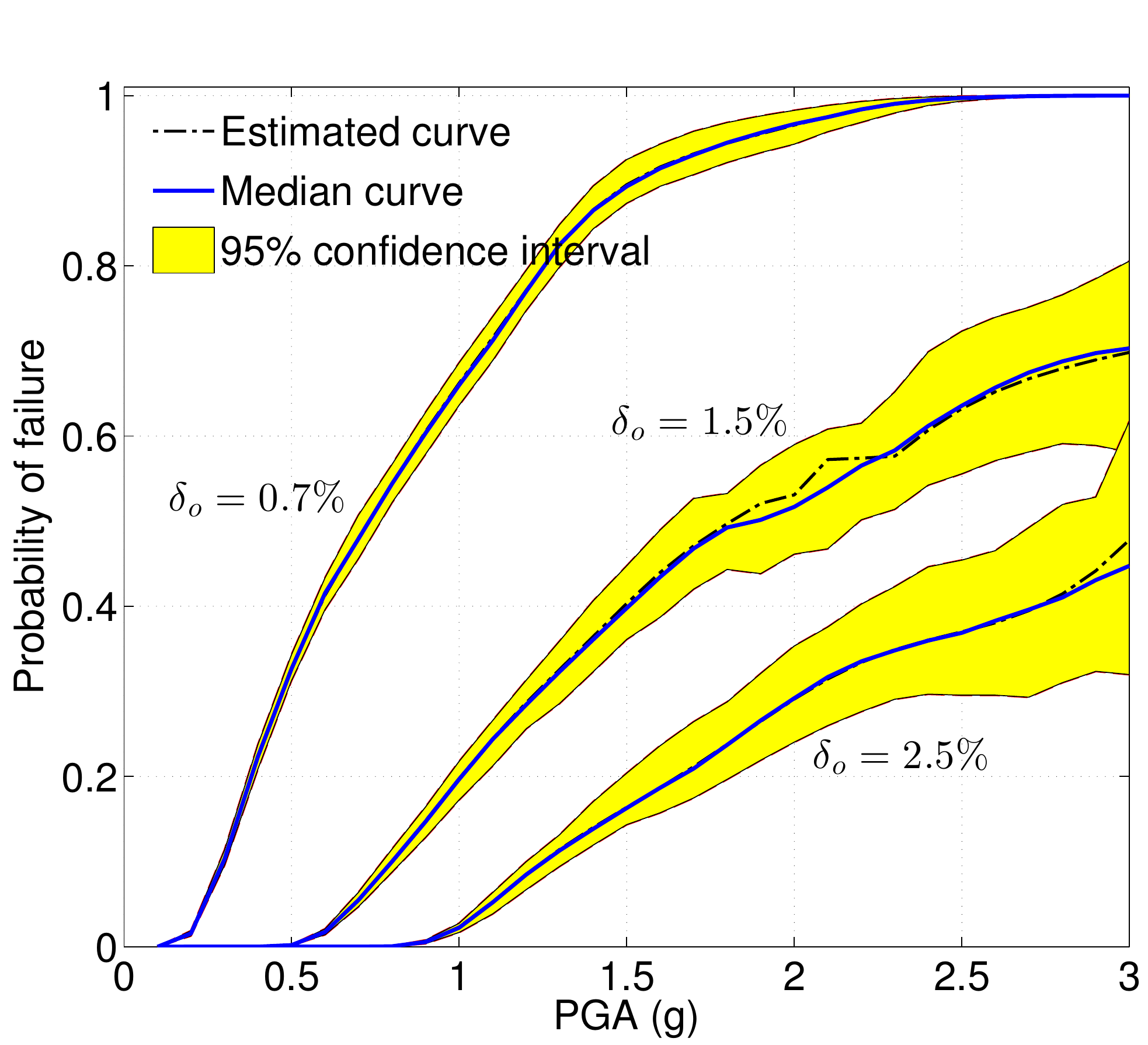}
	}
\subfigure
	{
		\includegraphics[width=0.45\textwidth]{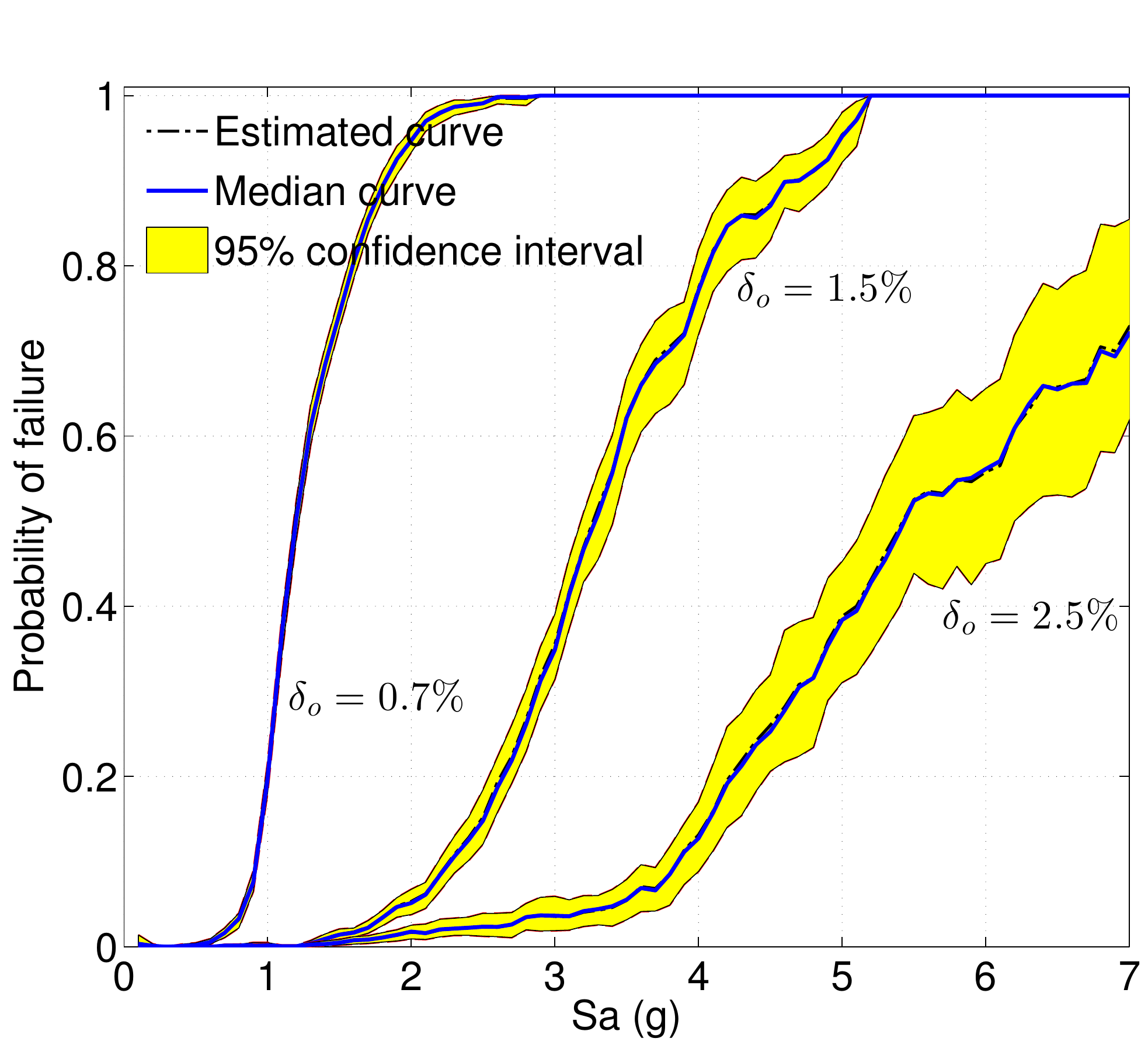}
	}
\subfigure
	{
		\includegraphics[width=0.45\textwidth]{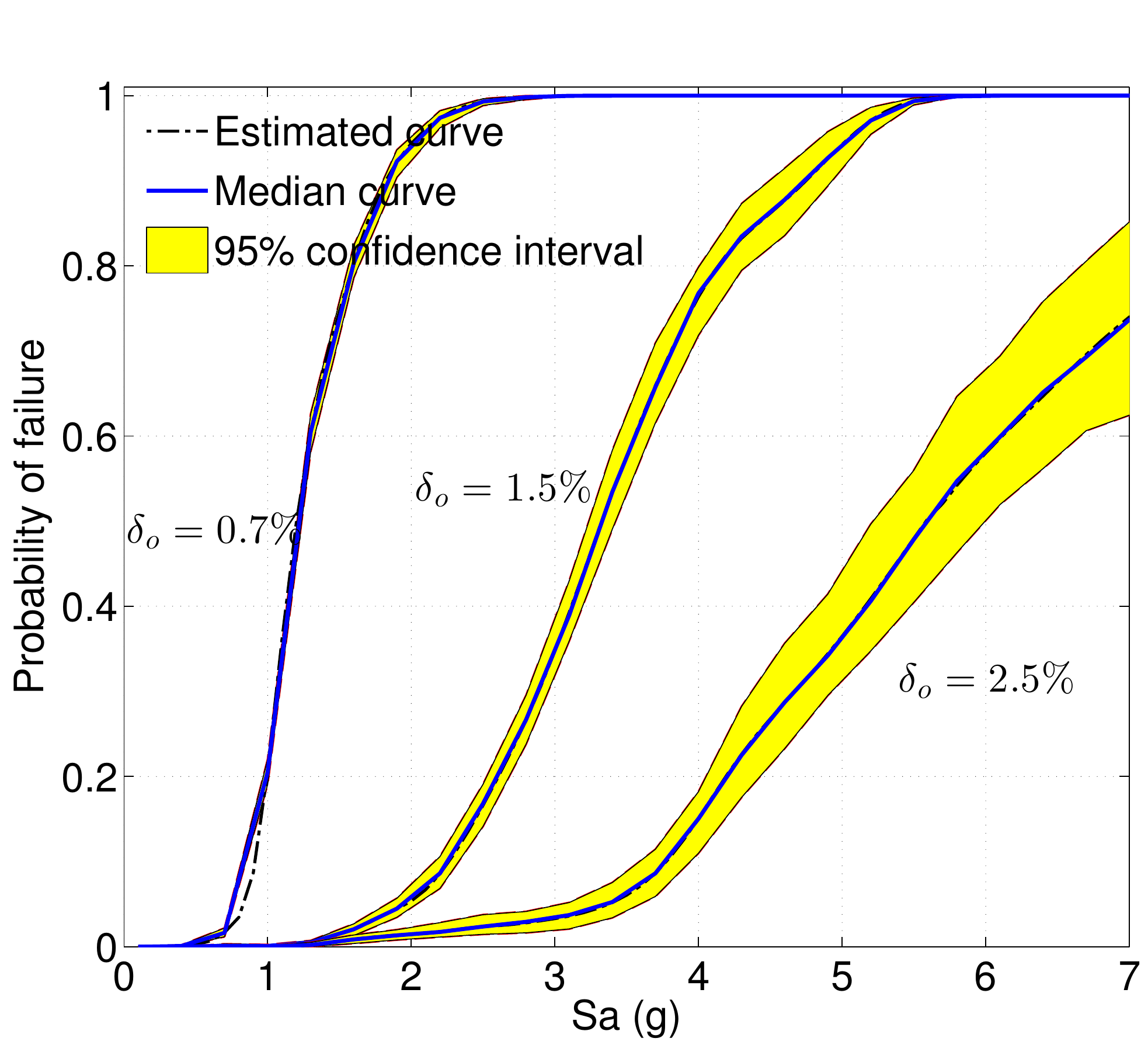}
	}
\caption{Estimated, mean bootstrap fragility curves and 95\% confidence intervals by binned Monte Carlo simulation and kernel density estimation approaches. The mean curves and 95\% confidence intervals are computed
 using bootstrap resampling technique with 100 replications.}
\label{figend}
\end{figure}

First, we examine the stability of the estimates by comparing those with the bootstrap medians, and the variability in the estimation by computing bootstrap confidence intervals. 
For the two considered $IM$s and the three limit states of interest, \figref{figend} shows the median bMCS- and KDE-based fragility curves and the 95\% confidence intervals obtained by bootstrap resampling with 100 replications.
\figref{figend} clearly shows that both the bMCS-based and the KDE-based median fragility curves obtained with the bootstrap method do not differ from the curves estimated with the original set of observations. This shows the stability of the proposed approaches. For a specified $IM$ and drift limit, the confidence intervals of the bMCS- and KDE-based curves have similar widths. The interval widths tend to increase with increasing drift limit and increasing $IM$ value. We note that for a certain drift limit and failure probability, the confidence intervals for the curves versus $Sa$ tend to be narrower than for curves versus $PGA$. This is due to the higher correlation of structural responses with $Sa$. 

In order to quantify the effect of epistemic uncertainty, one can assess the variability of the median $IM$, \ie the $IM$ value leading to 50\% probability of exceedance. Assuming that the median $IM$ ($PGA$ or $Sa$) follows a lognormal distribution \cite{Choun2010}, the median $IM$ is determined for each bootstrap curve and the log-standard deviation of the distribution of the median is computed. Table~\ref{tab:3} lists the log-standard deviations of the median $IM$ values for the same cases as in \figref{figend}. For the larger threshold  $\delta_o=2.5\%$ and $PGA$ as $IM$, some curves did not reach probability of exceedance values as high as 50\% and thus, results for this case are not shown. One observes in Table~\ref{tab:3} that epistemic uncertainty is increasing with increasing threshold $\delta_o$. For a certain threshold, using $PGA$ as $IM$ leads to a 4-5 times larger epistemic uncertainty compared to using $Sa$. Furthermore, log-standard deviations of the median $IM$s are slightly smaller with the KDE approach than with the bMCS approach. However, in all cases, the log-standard deviations are small indicating a low level of epistemic uncertainty. This is due to the large number of ground motions ($N=10^4)$ considered in this study.

\begin{table}[!ht]
\caption{Log-standard deviation of the median $IM$}
\centering
\begin{tabular}{cccccc}
\toprule
$\delta_o$ & Approach & $PGA$ & $Sa$  \\
\midrule  
\multirow{2}{*}{0.7\%}  
  & bMCS & $0.0289$ &  $0.0073$  \\
  & KDE & $0.0259$ & $0.0052$  \\

\midrule  
\multirow{2}{*}{1.5\%}  
  & bMCS & $0.0658$ &  $0.0168$  \\
  & KDE & $0.0639$ &  $0.0139$   \\
  
\midrule  
\multirow{2}{*}{2.5\%}  
  & bMCS & $$ &  $0.0493$  \\
  & KDE & $$ &  $0.0306$   \\

\bottomrule
\end{tabular}
\label{tab:3}
\end{table}

\section{Conclusions}

Seismic demand fragility evaluation is one of the basic elements in the framework of
performance-based earthquake engineering (PBEE). At present, the classical lognormal approach
is widely used to establish such fragility curves mainly due to the fact that
the lognormality assumption makes seismic risk analysis more tractable. The approach consists in
assigning the shape of a lognormal cumulative distribution function to the fragility 
curves. However, the validity of this assumption remains an open question.

In this paper, we introduce two non-parametric approaches in order to examine the validity of
the classical lognormal approach, namely the binned Monte Carlo simulation and the 
kernel density estimation. The former computes the crude Monte Carlo estimators 
for small subsets of ground motions with similar values of a selected intensity measure, while the latter 
estimates the conditional probability density function of the structural response 
given the ground motion intensity measures using the kernel density estimation 
technique. The proposed approaches can be used to compute fragility curves when the actual shape of these curves is not known as well as to validate or calibrate parametric fragility curves. Herein, the two non-parametric approaches are confronted to the classical 
lognormal approach on the example of a steel frame subject to synthetic ground motions.

In the case study, the fragility curves are established for three inter-storey drift-limit levels and two ground motion intensity 
measures, namely the peak ground acceleration ($PGA$) and the spectral acceleration ($Sa$).
The two non-parametric curves are consistent in all cases which proves the validity of the proposed techniques. 
Accordingly, these are used as reference to assess the accuracy of the lognormal curves. 
The parameters of the latter are estimated with two approaches, namely by maximum likelihood estimation and by assuming a linear probabilistic seismic demand model in the log-scale. 
For the two higher drift-limit levels, large discrepancies are observed between the non-parametric and the lognormal curves as well as between the two lognormal curves, which demonstrates the insufficiency of the parametric approach. Fragility estimates obtained with the latter may be conservative or not depending on the estimation approach, the intensity measure and the drift threshold.
When integrated in the PBEE framework, the discrepancies of the lognormal curves from the ``real" non-parametric ones may induce severe errors in the probabilistic consequence estimates that serve as decision variables for risk mitigation actions.

In the same case study, the bootstrap resampling technique is employed to assess effects of epistemic uncertainty in the non-parametric fragility curves. It is shown that for a certain limit state, epistemic uncertainty is lower for the case of $Sa$, which is a structure-specific intensity measure, than for the case of $PGA$. In addition, results from bootstrap analysis validate the stability of the fragility estimates with the proposed non-parametric methods.

Recently, fragility surfaces have emerged as
an innovative way to represent the system's vulnerability \cite{Seyedi2010} in which one calculates
the failure probability conditional on two intensity measures of the earthquake motions. The computation of these
surfaces is not straightforward and requires large computational effort.
The present study opens new paths for establishing the fragility surfaces:
similarly to the case of fragility curves, one can use kernel density estimation 
to obtain assumption-free fragility surfaces that are consistent with the true ones obtained by Monte Carlo simulation.

The computational cost of the two proposed approaches remains significant since they 
are based on rather large Monte Carlo samples. In order to reduce this cost, alternative 
approaches may be envisaged. 
Polynomial chaos (PC) expansions \cite{Ghanembook2003, BlatmanJCP2011} 
appear as a promising tool. Based on a smaller sample set (typically a few hundreds of finite element runs), PC 
expansion provides a polynomial approximation that surrogates the structural response. The 
feasibility of post-processing PC expansions in order to compute fragility curves has been 
shown in~\cite{SudretGuyonnet2011, SudretMaiIcossar2013} in the case a linear structural 
behavior is assumed. The extension to nonlinear behavior is currently in progress.

We underline that the proposed non-parametric approaches are essentially applicable to other probabilistic models in the PBEE framework, relating decision variables with structural damage and structural damage with structural response.  Once all the non-parametric probabilistic models are available, they can be incorporated in the PBEE framework by means of numerical integration. Then a full seismic risk assessment may be conducted by avoiding potential inaccuracies introduced from simplifying parametric assumptions at any step of the analysis. Optimal high-fidelity computational methods for incorporating non-parametric fragility curves in the PBEE framework will be investigated in the future.





\bibliographystyle{model3-num-names} 
\bibliography{EESD2014}

\begin{thebibliography}{64}
\providecommand{\natexlab}[1]{#1}
\providecommand{\url}[1]{\texttt{#1}}
\providecommand{\href}[2]{#2}
\providecommand{\path}[1]{#1}
\providecommand{\eprint}[1]{\href{http://arxiv.org/abs/#1}{\path{#1}}}
\providecommand{\DOIprefix}{doi:}
\providecommand{\ArXivprefix}{arXiv:}
\providecommand{\URLprefix}{URL: }
\providecommand{\Pubmedprefix}{pmid:}
\providecommand{\doi}[1]{\href{http://dx.doi.org/#1}{\path{#1}}}
\providecommand{\Pubmed}[1]{\href{pmid:#1}{\path{#1}}}
\providecommand{\BIBand}{and}
\providecommand{\bibinfo}[2]{#2}
\ifx\xfnm\undefined \def\xfnm[#1]{\unskip,\space#1}\fi
\bibitem[{Porter(2003)}]{Porter2003}
\bibinfo{author}{Porter\xfnm[ K.A.]}.
\newblock \bibinfo{title}{{An overview of PEER's performance-based earthquake
  engineering methodology}}.
\newblock In: \bibinfo{booktitle}{Proc. 9th Int. Conf. on Applications of Stat.
  and Prob. in Civil Engineering (ICASP9), San Francisco}.
  \bibinfo{year}{2003}, p. \bibinfo{pages}{6--9}.
\bibitem[{Baker and Cornell(2008)}]{Baker2008a}
\bibinfo{author}{Baker\xfnm[ J.W.]}, \bibinfo{author}{Cornell\xfnm[ C.A.]}.
\newblock \bibinfo{title}{{Uncertainty propagation in probabilistic seismic
  loss estimation}}.
\newblock \bibinfo{journal}{Structural Safety}
  \bibinfo{year}{2008};\bibinfo{volume}{30}(\bibinfo{number}{3}):\bibinfo{pages}{236--252}.
\bibitem[{G\"{u}nay and Mosalam(2013)}]{Gunay2013}
\bibinfo{author}{G\"{u}nay\xfnm[ S.]}, \bibinfo{author}{Mosalam\xfnm[ K.M.]}.
\newblock \bibinfo{title}{{PEER performance-based earthquake engineering
  methodology, revisited}}.
\newblock \bibinfo{journal}{J Earthq Eng}
  \bibinfo{year}{2013};\bibinfo{volume}{17}(\bibinfo{number}{6}):\bibinfo{pages}{829--858}.
\bibitem[{Mackie and Stojadinovic(2005)}]{Mackie2005}
\bibinfo{author}{Mackie\xfnm[ K.]}, \bibinfo{author}{Stojadinovic\xfnm[ B.]}.
\newblock \bibinfo{title}{{Fragility basis for California highway overpass
  bridge seismic decision making}}.
\newblock \bibinfo{publisher}{Pacific Earthquake Engineering Research Center,
  College of Engineering, University of California, Berkeley};
  \bibinfo{year}{2005}.
\bibitem[{Ellingwood and Kinali(2009)}]{Ellingwood2009}
\bibinfo{author}{Ellingwood\xfnm[ B.R.]}, \bibinfo{author}{Kinali\xfnm[ K.]}.
\newblock \bibinfo{title}{{Quantifying and communicating uncertainty in seismic
  risk assessment}}.
\newblock \bibinfo{journal}{Structural Safety}
  \bibinfo{year}{2009};\bibinfo{volume}{31}(\bibinfo{number}{2}):\bibinfo{pages}{179--187}.
\bibitem[{Seo et~al.(2012)Seo, Duenas-Osorio, Craig and Goodno}]{Seo2012}
\bibinfo{author}{Seo\xfnm[ J.]}, \bibinfo{author}{Duenas-Osorio\xfnm[ L.]},
  \bibinfo{author}{Craig\xfnm[ J.I.]}, \bibinfo{author}{Goodno\xfnm[ B.J.]}.
\newblock \bibinfo{title}{{Metamodel-based regional vulnerability estimate of
  irregular steel moment-frame structures subjected to earthquake events}}.
\newblock \bibinfo{journal}{Eng Struct}
  \bibinfo{year}{2012};\bibinfo{volume}{45}:\bibinfo{pages}{585--597}.
\bibitem[{Banerjee and Shinozuka(2007)}]{Banerjee2007}
\bibinfo{author}{Banerjee\xfnm[ S.]}, \bibinfo{author}{Shinozuka\xfnm[ M.]}.
\newblock \bibinfo{title}{{Nonlinear Static Procedure for Seismic Vulnerability
  Assessment of Bridges}}.
\newblock \bibinfo{journal}{Comput-Aided Civ Inf}
  \bibinfo{year}{2007};\bibinfo{volume}{22}(\bibinfo{number}{4}):\bibinfo{pages}{293--305}.
\bibitem[{Richardson et~al.(1980)Richardson, Bagchi and
  Brazee}]{Richardson1980}
\bibinfo{author}{Richardson\xfnm[ J.E.]}, \bibinfo{author}{Bagchi\xfnm[ G.]},
  \bibinfo{author}{Brazee\xfnm[ R.J.]}.
\newblock \bibinfo{title}{{The seismic safety margins research program of the
  U.S. Nuclear Regulatory Commission}}.
\newblock \bibinfo{journal}{Nuc Eng Des}
  \bibinfo{year}{1980};\bibinfo{volume}{59}(\bibinfo{number}{1}):\bibinfo{pages}{15--25}.
\bibitem[{Pei and {Van De Lindt}(2009)}]{Pei2009}
\bibinfo{author}{Pei\xfnm[ S.]}, \bibinfo{author}{{Van De Lindt}\xfnm[ J.]}.
\newblock \bibinfo{title}{{Methodology for earthquake-induced loss estimation:
  An application to woodframe buildings}}.
\newblock \bibinfo{journal}{Structural Safety}
  \bibinfo{year}{2009};\bibinfo{volume}{31}(\bibinfo{number}{1}):\bibinfo{pages}{31--42}.
\bibitem[{Eads et~al.(2013)Eads, Miranda, Krawinkler and Lignos}]{Eads2013}
\bibinfo{author}{Eads\xfnm[ L.]}, \bibinfo{author}{Miranda\xfnm[ E.]},
  \bibinfo{author}{Krawinkler\xfnm[ H.]}, \bibinfo{author}{Lignos\xfnm[ D.G.]}.
\newblock \bibinfo{title}{{An efficient method for estimating the collapse risk
  of structures in seismic regions}}.
\newblock \bibinfo{journal}{Earthquake Eng Struct Dyn}
  \bibinfo{year}{2013};\bibinfo{volume}{42}(\bibinfo{number}{1}):\bibinfo{pages}{25--41}.
\bibitem[{Dukes et~al.(2012)Dukes, DesRoches and Padgett}]{Dukes2012}
\bibinfo{author}{Dukes\xfnm[ J.]}, \bibinfo{author}{DesRoches\xfnm[ R.]},
  \bibinfo{author}{Padgett\xfnm[ J.E.]}.
\newblock \bibinfo{title}{{Sensitivity study of design parameters used to
  develop bridge specific fragility curves}}.
\newblock In: \bibinfo{booktitle}{Proc. 15th World Conf. Earthquake Eng.}
  \bibinfo{year}{2012},.
\bibitem[{G\"{u}neyisi and Altay(2008)}]{Guneyisi2008}
\bibinfo{author}{G\"{u}neyisi\xfnm[ E.M.]}, \bibinfo{author}{Altay\xfnm[ G.]}.
\newblock \bibinfo{title}{{Seismic fragility assessment of effectiveness of
  viscous dampers in R/C buildings under scenario earthquakes}}.
\newblock \bibinfo{journal}{Structural Safety}
  \bibinfo{year}{2008};\bibinfo{volume}{30}(\bibinfo{number}{5}):\bibinfo{pages}{461--480}.
\bibitem[{Seyedi et~al.(2010)Seyedi, Gehl, Douglas, Davenne, Mezher and
  Ghavamian}]{Seyedi2010}
\bibinfo{author}{Seyedi\xfnm[ D.M.]}, \bibinfo{author}{Gehl\xfnm[ P.]},
  \bibinfo{author}{Douglas\xfnm[ J.]}, \bibinfo{author}{Davenne\xfnm[ L.]},
  \bibinfo{author}{Mezher\xfnm[ N.]}, \bibinfo{author}{Ghavamian\xfnm[ S.]}.
\newblock \bibinfo{title}{{Development of seismic fragility surfaces for
  reinforced concrete buildings by means of nonlinear time-history analysis}}.
\newblock \bibinfo{journal}{Earthquake Eng Struct Dyn}
  \bibinfo{year}{2010};\bibinfo{volume}{39}(\bibinfo{number}{1}):\bibinfo{pages}{91--108}.
\bibitem[{Gardoni et~al.(2002)Gardoni, {Der Kiureghian} and
  Mosalam}]{Gardoni2002a}
\bibinfo{author}{Gardoni\xfnm[ P.]}, \bibinfo{author}{{Der Kiureghian}\xfnm[
  A.]}, \bibinfo{author}{Mosalam\xfnm[ K.M.]}.
\newblock \bibinfo{title}{{Probabilistic capacity models and fragility
  estimates for reinforced concrete columns based on experimental
  observations}}.
\newblock \bibinfo{journal}{J Eng Mech}
  \bibinfo{year}{2002};\bibinfo{volume}{128}(\bibinfo{number}{10}):\bibinfo{pages}{1024--1038}.
\bibitem[{Ghosh and Padgett(2010)}]{Ghosh2010}
\bibinfo{author}{Ghosh\xfnm[ J.]}, \bibinfo{author}{Padgett\xfnm[ J.E.]}.
\newblock \bibinfo{title}{{Aging considerations in the development of
  time-dependent seismic fragility curves}}.
\newblock \bibinfo{journal}{J Struct Eng}
  \bibinfo{year}{2010};\bibinfo{volume}{136}(\bibinfo{number}{12}):\bibinfo{pages}{1497--1511}.
\bibitem[{Argyroudis and Pitilakis(2012)}]{Argyroudis2012}
\bibinfo{author}{Argyroudis\xfnm[ S.]}, \bibinfo{author}{Pitilakis\xfnm[ K.]}.
\newblock \bibinfo{title}{{Seismic fragility curves of shallow tunnels in
  alluvial deposits}}.
\newblock \bibinfo{journal}{Soil Dyn Earthq Eng}
  \bibinfo{year}{2012};\bibinfo{volume}{35}:\bibinfo{pages}{1--12}.
\bibitem[{Chiou et~al.(2011)Chiou, Chiang, Yang and Hsu}]{Chiou2011}
\bibinfo{author}{Chiou\xfnm[ J.]}, \bibinfo{author}{Chiang\xfnm[ C.]},
  \bibinfo{author}{Yang\xfnm[ H.]}, \bibinfo{author}{Hsu\xfnm[ S.]}.
\newblock \bibinfo{title}{{Developing fragility curves for a pile-supported
  wharf}}.
\newblock \bibinfo{journal}{Soil Dyn Earthq Eng}
  \bibinfo{year}{2011};\bibinfo{volume}{31}:\bibinfo{pages}{830--840}.
\bibitem[{Quilligan et~al.(2012)Quilligan, {O Connor} and
  Pakrashi}]{Quilligan2012}
\bibinfo{author}{Quilligan\xfnm[ A.]}, \bibinfo{author}{{O Connor}\xfnm[ A.]},
  \bibinfo{author}{Pakrashi\xfnm[ V.]}.
\newblock \bibinfo{title}{{Fragility analysis of steel and concrete wind
  turbine towers}}.
\newblock \bibinfo{journal}{Engineering Structures}
  \bibinfo{year}{2012};\bibinfo{volume}{36}:\bibinfo{pages}{270--282}.
\bibitem[{Borgonovo et~al.(2013)Borgonovo, Zentner, Pellegri, Tarantola and
  de~Rocquigny}]{Borgonovo2013}
\bibinfo{author}{Borgonovo\xfnm[ E.]}, \bibinfo{author}{Zentner\xfnm[ I.]},
  \bibinfo{author}{Pellegri\xfnm[ A.]}, \bibinfo{author}{Tarantola\xfnm[ S.]},
  \bibinfo{author}{de~Rocquigny\xfnm[ E.]}.
\newblock \bibinfo{title}{{On the importance of uncertain factors in seismic
  fragility assessment}}.
\newblock \bibinfo{journal}{Reliab Eng Sys Safety}
  \bibinfo{year}{2013};\bibinfo{volume}{109}(\bibinfo{number}{0}):\bibinfo{pages}{66--76}.
\bibitem[{Rossetto and Elnashai(2005)}]{Rossetto2005}
\bibinfo{author}{Rossetto\xfnm[ T.]}, \bibinfo{author}{Elnashai\xfnm[ A.]}.
\newblock \bibinfo{title}{{A new analytical procedure for the derivation of
  displacement-based vulnerability curves for populations of RC structures}}.
\newblock \bibinfo{journal}{Engineering Structures}
  \bibinfo{year}{2005};\bibinfo{volume}{27}(\bibinfo{number}{3}):\bibinfo{pages}{397--409}.
\bibitem[{Shinozuka et~al.(2000)Shinozuka, Feng, Lee and
  Naganuma}]{Shinozuka2000b}
\bibinfo{author}{Shinozuka\xfnm[ M.]}, \bibinfo{author}{Feng\xfnm[ M.]},
  \bibinfo{author}{Lee\xfnm[ J.]}, \bibinfo{author}{Naganuma\xfnm[ T.]}.
\newblock \bibinfo{title}{{Statistical analysis of fragility curves}}.
\newblock \bibinfo{journal}{J Eng Mech}
  \bibinfo{year}{2000};\bibinfo{volume}{126}(\bibinfo{number}{12}):\bibinfo{pages}{1224--1231}.
\bibitem[{Ellingwood(2001)}]{Ellingwood2001}
\bibinfo{author}{Ellingwood\xfnm[ B.R.]}.
\newblock \bibinfo{title}{{Earthquake risk assessment of building structures}}.
\newblock \bibinfo{journal}{Reliab Eng Sys Safety}
  \bibinfo{year}{2001};\bibinfo{volume}{74}(\bibinfo{number}{3}):\bibinfo{pages}{251--262}.
\bibitem[{Zentner(2010)}]{Zentner2010a}
\bibinfo{author}{Zentner\xfnm[ I.]}.
\newblock \bibinfo{title}{{Numerical computation of fragility curves for NPP
  equipment}}.
\newblock \bibinfo{journal}{Nuc Eng Des}
  \bibinfo{year}{2010};\bibinfo{volume}{240}:\bibinfo{pages}{1614--1621}.
\bibitem[{Gencturk et~al.(2008)Gencturk, Elnashai and Song}]{Gencturk2008}
\bibinfo{author}{Gencturk\xfnm[ B.]}, \bibinfo{author}{Elnashai\xfnm[ A.]},
  \bibinfo{author}{Song\xfnm[ J.]}.
\newblock \bibinfo{title}{{Fragility relationships for populations of woodframe
  structures based on inelastic response}}.
\newblock \bibinfo{journal}{J Earthq Eng}
  \bibinfo{year}{2008};\bibinfo{volume}{12}:\bibinfo{pages}{119--128}.
\bibitem[{Jeong et~al.(2012)Jeong, Mwafy and Elnashai}]{Jeong2012}
\bibinfo{author}{Jeong\xfnm[ S.H.]}, \bibinfo{author}{Mwafy\xfnm[ A.M.]},
  \bibinfo{author}{Elnashai\xfnm[ A.S.]}.
\newblock \bibinfo{title}{{Probabilistic seismic performance assessment of
  code-compliant multi-story RC buildings}}.
\newblock \bibinfo{journal}{Engineering Structures}
  \bibinfo{year}{2012};\bibinfo{volume}{34}:\bibinfo{pages}{527--537}.
\bibitem[{Banerjee and Shinozuka(2008)}]{Banerjee2008}
\bibinfo{author}{Banerjee\xfnm[ S.]}, \bibinfo{author}{Shinozuka\xfnm[ M.]}.
\newblock \bibinfo{title}{{Mechanistic quantification of RC bridge damage
  states under earthquake through fragility analysis}}.
\newblock \bibinfo{journal}{Prob Eng Mech}
  \bibinfo{year}{2008};\bibinfo{volume}{23}(\bibinfo{number}{1}):\bibinfo{pages}{12--22}.
\bibitem[{Hwang and Huo(1994)}]{Hwang1994}
\bibinfo{author}{Hwang\xfnm[ H.H.]}, \bibinfo{author}{Huo\xfnm[ J.R.]}.
\newblock \bibinfo{title}{{Generation of hazard-consistent fragility curves}}.
\newblock \bibinfo{journal}{Soil Dyn Earthq Eng}
  \bibinfo{year}{1994};\bibinfo{volume}{13}(\bibinfo{number}{5}):\bibinfo{pages}{345--354}.
\bibitem[{Lupoi et~al.(2005)Lupoi, Franchin, Pinto and Monti}]{Lupoi2005}
\bibinfo{author}{Lupoi\xfnm[ A.]}, \bibinfo{author}{Franchin\xfnm[ P.]},
  \bibinfo{author}{Pinto\xfnm[ P.E.]}, \bibinfo{author}{Monti\xfnm[ G.]}.
\newblock \bibinfo{title}{{Seismic design of bridges accounting for spatial
  variability of ground motion}}.
\newblock \bibinfo{journal}{Earthquake Eng Struct Dyn}
  \bibinfo{year}{2005};\bibinfo{volume}{34}(\bibinfo{number}{4-5}):\bibinfo{pages}{327--348}.
\bibitem[{Choi et~al.(2004)Choi, DesRoches and Nielson}]{Choi2004}
\bibinfo{author}{Choi\xfnm[ E.]}, \bibinfo{author}{DesRoches\xfnm[ R.]},
  \bibinfo{author}{Nielson\xfnm[ B.]}.
\newblock \bibinfo{title}{{Seismic fragility of typical bridges in moderate
  seismic zones}}.
\newblock \bibinfo{journal}{Eng Struct}
  \bibinfo{year}{2004};\bibinfo{volume}{26}(\bibinfo{number}{2}):\bibinfo{pages}{187--199}.
\bibitem[{Rezaeian and {Der Kiureghian}(2008)}]{Rezaeian2008}
\bibinfo{author}{Rezaeian\xfnm[ S.]}, \bibinfo{author}{{Der Kiureghian}\xfnm[
  A.]}.
\newblock \bibinfo{title}{{A stochastic ground motion model with separable
  temporal and spectral nonstationarities}}.
\newblock \bibinfo{journal}{Earthquake Eng Struct Dyn}
  \bibinfo{year}{2008};\bibinfo{volume}{37}(\bibinfo{number}{13}):\bibinfo{pages}{1565--1584}.
\bibitem[{Rezaeian and {Der Kiureghian}(2010)}]{Rezaeian2010}
\bibinfo{author}{Rezaeian\xfnm[ S.]}, \bibinfo{author}{{Der Kiureghian}\xfnm[
  A.]}.
\newblock \bibinfo{title}{{Simulation of synthetic ground motions for specified
  earthquake and site characteristics}}.
\newblock \bibinfo{journal}{Earthquake Eng Struct Dyn}
  \bibinfo{year}{2010};\bibinfo{volume}{39}(\bibinfo{number}{10}):\bibinfo{pages}{1155--1180}.
\bibitem[{Padgett and DesRoches(2008)}]{Padgett2008}
\bibinfo{author}{Padgett\xfnm[ J.E.]}, \bibinfo{author}{DesRoches\xfnm[ R.]}.
\newblock \bibinfo{title}{{Methodology for the development of analytical
  fragility curves for retrofitted bridges}}.
\newblock \bibinfo{journal}{Earthquake Eng Struct Dyn}
  \bibinfo{year}{2008};\bibinfo{volume}{37}(\bibinfo{number}{8}):\bibinfo{pages}{1157--1174}.
\bibitem[{Zareian and Krawinkler(2007)}]{Zareian2007}
\bibinfo{author}{Zareian\xfnm[ F.]}, \bibinfo{author}{Krawinkler\xfnm[ H.]}.
\newblock \bibinfo{title}{{Assessment of probability of collapse and design for
  collapse safety}}.
\newblock \bibinfo{journal}{Earthquake Eng Struct Dyn}
  \bibinfo{year}{2007};\bibinfo{volume}{36}(\bibinfo{number}{13}):\bibinfo{pages}{1901--1914}.
\bibitem[{Vamvatsikos and Cornell(2002)}]{Vamvatsikos2002}
\bibinfo{author}{Vamvatsikos\xfnm[ D.]}, \bibinfo{author}{Cornell\xfnm[ C.A.]}.
\newblock \bibinfo{title}{{Incremental dynamic analysis}}.
\newblock \bibinfo{journal}{Earthquake Eng Struct Dyn}
  \bibinfo{year}{2002};\bibinfo{volume}{31}(\bibinfo{number}{3}):\bibinfo{pages}{491--514}.
\bibitem[{Vamvatsikos and Cornell(2005)}]{Vamvatsikos2005}
\bibinfo{author}{Vamvatsikos\xfnm[ D.]}, \bibinfo{author}{Cornell\xfnm[ C.A.]}.
\newblock \bibinfo{title}{{Direct estimation of seismic demand and capacity of
  multidegree-of-freedom systems through Iincremental dynamic analysis of
  single degree of freedom approximation1}}.
\newblock \bibinfo{journal}{J Struct Eng}
  \bibinfo{year}{2005};\bibinfo{volume}{131}(\bibinfo{number}{4}):\bibinfo{pages}{589--599}.
\bibitem[{Mander et~al.(2007)Mander, Dhakal, Mashiko and Solberg}]{Mander2007}
\bibinfo{author}{Mander\xfnm[ J.B.]}, \bibinfo{author}{Dhakal\xfnm[ R.P.]},
  \bibinfo{author}{Mashiko\xfnm[ N.]}, \bibinfo{author}{Solberg\xfnm[ K.M.]}.
\newblock \bibinfo{title}{{Incremental dynamic analysis applied to seismic
  financial risk assessment of bridges}}.
\newblock \bibinfo{journal}{Eng Struct}
  \bibinfo{year}{2007};\bibinfo{volume}{29}(\bibinfo{number}{10}):\bibinfo{pages}{2662--2672}.
\bibitem[{Mehdizadeh et~al.(2012)Mehdizadeh, Mackie and
  Nielson}]{Mehdizadeh2012}
\bibinfo{author}{Mehdizadeh\xfnm[ M.]}, \bibinfo{author}{Mackie\xfnm[ K.R.]},
  \bibinfo{author}{Nielson\xfnm[ B.G.]}.
\newblock \bibinfo{title}{{Scaling bias and record selection for fragility
  analysis}}.
\newblock In: \bibinfo{booktitle}{Proc. 15th World Conf. Earthquake Eng.}
  \bibinfo{year}{2012},.
\bibitem[{Cimellaro et~al.(2009)Cimellaro, Reinhorn, D'Ambrisi and {De
  Stefano}}]{Cimellaro2009}
\bibinfo{author}{Cimellaro\xfnm[ G.P.]}, \bibinfo{author}{Reinhorn\xfnm[
  A.M.]}, \bibinfo{author}{D'Ambrisi\xfnm[ A.]}, \bibinfo{author}{{De
  Stefano}\xfnm[ M.]}.
\newblock \bibinfo{title}{{Fragility analysis and seismic record selection}}.
\newblock \bibinfo{journal}{J Struct Eng}
  \bibinfo{year}{2009};\bibinfo{volume}{137}(\bibinfo{number}{3}):\bibinfo{pages}{379--390}.
\bibitem[{Wand and Jones(1995)}]{WandJones}
\bibinfo{author}{Wand\xfnm[ M.]}, \bibinfo{author}{Jones\xfnm[ M.C.]}.
\newblock \bibinfo{title}{{Kernel smoothing}}.
\newblock \bibinfo{publisher}{Chapman and Hall}; \bibinfo{year}{1995}.
\bibitem[{Duong(2004)}]{DuongThesis2004}
\bibinfo{author}{Duong\xfnm[ T.]}.
\newblock \bibinfo{title}{{Bandwidth selectors for multivariate kernel density
  estimation}}.
\newblock Ph.D. thesis; School of mathematics and Statistics, University of
  Western Australia; \bibinfo{year}{2004}.
\bibitem[{Duong and Hazelton(2005)}]{Duong2005}
\bibinfo{author}{Duong\xfnm[ T.]}, \bibinfo{author}{Hazelton\xfnm[ M.L.]}.
\newblock \bibinfo{title}{{Cross-validation bandwidth matrices for multivariate
  kernel density estimation}}.
\newblock \bibinfo{journal}{Scand J Stat}
  \bibinfo{year}{2005};\bibinfo{volume}{32}(\bibinfo{number}{3}):\bibinfo{pages}{485--506}.
\bibitem[{Frankel et~al.(2000)Frankel, Mueller, Barnhard, Leyendecker, Wesson,
  Harmsen et~al.}]{Frankel2000}
\bibinfo{author}{Frankel\xfnm[ A.D.]}, \bibinfo{author}{Mueller\xfnm[ C.S.]},
  \bibinfo{author}{Barnhard\xfnm[ T.P.]}, \bibinfo{author}{Leyendecker\xfnm[
  E.V.]}, \bibinfo{author}{Wesson\xfnm[ R.L.]}, \bibinfo{author}{Harmsen\xfnm[
  S.C.]}, et~al.
\newblock \bibinfo{title}{{USGS national seismic hazard maps}}.
\newblock \bibinfo{journal}{Earthquake spectra}
  \bibinfo{year}{2000};\bibinfo{volume}{16}(\bibinfo{number}{1}):\bibinfo{pages}{1--19}.
\bibitem[{Sudret and Mai(2013{\natexlab{a}})}]{SudretMaiCFM2013}
\bibinfo{author}{Sudret\xfnm[ B.]}, \bibinfo{author}{Mai\xfnm[ C.V.]}.
\newblock \bibinfo{title}{{Calcul des courbes de fragilit\'{e} par approches
  non-param\'{e}triques}}.
\newblock In: \bibinfo{booktitle}{Proc. 21e Congr\`{e}s Fran\c{c}ais de
  M\'{e}canique (CFM21), Bordeaux}. \bibinfo{year}{2013}{\natexlab{a}},.
\bibitem[{Bradley and Lee(2010)}]{Bradley2010}
\bibinfo{author}{Bradley\xfnm[ B.A.]}, \bibinfo{author}{Lee\xfnm[ D.S.]}.
\newblock \bibinfo{title}{{Accuracy of approximate methods of uncertainty
  propagation in seismic loss estimation}}.
\newblock \bibinfo{journal}{Structural Safety}
  \bibinfo{year}{2010};\bibinfo{volume}{32}(\bibinfo{number}{1}):\bibinfo{pages}{13--24}.
\bibitem[{Liel et~al.(2009)Liel, Haselton, Deierlein and Baker}]{Liel2009}
\bibinfo{author}{Liel\xfnm[ A.B.]}, \bibinfo{author}{Haselton\xfnm[ C.B.]},
  \bibinfo{author}{Deierlein\xfnm[ G.G.]}, \bibinfo{author}{Baker\xfnm[ J.W.]}.
\newblock \bibinfo{title}{{Incorporating modeling uncertainties in the
  assessment of seismic collapse risk of buildings}}.
\newblock \bibinfo{journal}{Structural Safety}
  \bibinfo{year}{2009};\bibinfo{volume}{31}(\bibinfo{number}{2}):\bibinfo{pages}{197--211}.
\bibitem[{Efron(1979)}]{Efron1979}
\bibinfo{author}{Efron\xfnm[ B.]}.
\newblock \bibinfo{title}{{Bootstrap methods: another look at the Jackknife}}.
\newblock \bibinfo{journal}{The annals of Statistics}
  \bibinfo{year}{1979};:\bibinfo{pages}{1--26}.
\bibitem[{{Pacific Earthquake Engineering and Research
  Center}(2004)}]{OpenSees}
\bibinfo{author}{{Pacific Earthquake Engineering and Research Center}\xfnm[]}.
\newblock \bibinfo{title}{{OpenSees: The Open System for Earthquake Engineering
  Simulation}}.
\newblock \bibinfo{year}{2004}.
\bibitem[{{Eurocode 3}(2005)}]{EC3}
\bibinfo{author}{{Eurocode 3}\xfnm[]}.
\newblock \bibinfo{title}{{Design of steel structures - Part 1-1: General rules
  and rules for buildings}}.
\newblock \bibinfo{year}{2005}.
\bibitem[{{Joint Committee on Structural Safety}(2001)}]{jcss}
\bibinfo{author}{{Joint Committee on Structural Safety}\xfnm[]}.
\newblock \bibinfo{title}{{Probabilistic Model Code - Part 3 : Resistance
  Models}}.
\newblock \bibinfo{year}{2001}.
\bibitem[{Deierlein et~al.(2010)Deierlein, Reinhorn and
  Willford}]{Deierlein2010}
\bibinfo{author}{Deierlein\xfnm[ G.G.]}, \bibinfo{author}{Reinhorn\xfnm[
  A.M.]}, \bibinfo{author}{Willford\xfnm[ M.R.]}.
\newblock \bibinfo{title}{{Nonlinear structural analysis for seismic design}}.
\newblock \bibinfo{journal}{NEHRP Seismic Design Technical Brief No}
  \bibinfo{year}{2010};\bibinfo{volume}{4}.
\bibitem[{{Eurocode 1}(2004)}]{EC1}
\bibinfo{author}{{Eurocode 1}\xfnm[]}.
\newblock \bibinfo{title}{{Actions on structures - Part 1-1: general actions -
  densities, self-weight, imposed loads for buildings}}.
\newblock \bibinfo{year}{2004}.
\bibitem[{Mackie and Stojadinovic(2004)}]{Mackie2004}
\bibinfo{author}{Mackie\xfnm[ K.]}, \bibinfo{author}{Stojadinovic\xfnm[ B.]}.
\newblock \bibinfo{title}{{Improving probabilistic seismic demand models
  through refined intensity measures}}.
\newblock In: \bibinfo{booktitle}{Proc. 13th World Conf. Earthquake Eng.}
  \bibinfo{publisher}{Int. Assoc. for Earthquake Eng. Japan};
  \bibinfo{year}{2004},.
\bibitem[{Padgett et~al.(2008)Padgett, Nielson and DesRoches}]{Padgett2008a}
\bibinfo{author}{Padgett\xfnm[ J.]}, \bibinfo{author}{Nielson\xfnm[ B.]},
  \bibinfo{author}{DesRoches\xfnm[ R.]}.
\newblock \bibinfo{title}{{Selection of optimal intensity measures in
  probabilistic seismic demand models of highway bridge portfolios}}.
\newblock \bibinfo{journal}{Earthquake Eng Struct Dyn}
  \bibinfo{year}{2008};\bibinfo{volume}{37}(\bibinfo{number}{5}):\bibinfo{pages}{711--725}.
\bibitem[{Cornell et~al.(2002)Cornell, Jalayer, Hamburger and
  Foutch}]{Cornell2002}
\bibinfo{author}{Cornell\xfnm[ C.]}, \bibinfo{author}{Jalayer\xfnm[ F.]},
  \bibinfo{author}{Hamburger\xfnm[ R.]}, \bibinfo{author}{Foutch\xfnm[ D.]}.
\newblock \bibinfo{title}{{Probabilistic basis for 2000 SAC Federal Emergency
  Management Agency steel moment frame guidelines}}.
\newblock \bibinfo{journal}{J Struct Eng (ASCE)}
  \bibinfo{year}{2002};\bibinfo{volume}{128}(\bibinfo{number}{4}):\bibinfo{pages}{526--533}.
\bibitem[{Lagaros and Fragiadakis(2007)}]{Lagaros2007}
\bibinfo{author}{Lagaros\xfnm[ N.D.]}, \bibinfo{author}{Fragiadakis\xfnm[ M.]}.
\newblock \bibinfo{title}{{Fragility sssessment of steel frames using neural
  networks}}.
\newblock \bibinfo{journal}{Earthquake Spectra}
  \bibinfo{year}{2007};\bibinfo{volume}{23}(\bibinfo{number}{4}):\bibinfo{pages}{735--752}.
\bibitem[{{Federal Emergency Management Agency, Washington,
  DC}(2000)}]{FEMA2000}
\bibinfo{author}{{Federal Emergency Management Agency, Washington, DC}\xfnm[]}.
\newblock \bibinfo{title}{{Commentary for the seismic rehabilitation of
  buildings}}; \bibinfo{year}{2000}.
\bibitem[{{Eurocode 8}(2004)}]{EC8eng}
\bibinfo{author}{{Eurocode 8}\xfnm[]}.
\newblock \bibinfo{title}{{Design of structures for earthquake resistance -
  Part 1: General rules, seismic actions and rules for buildings}};
  \bibinfo{year}{2004}.
\bibitem[{Duong(2007)}]{Duongks2007}
\bibinfo{author}{Duong\xfnm[ T.]}.
\newblock \bibinfo{title}{{ks: kernel density estimation and kernel
  discriminant analysis for multivariate data in R}}.
\newblock \bibinfo{journal}{J Stat Softw}
  \bibinfo{year}{2007};\bibinfo{volume}{21}(\bibinfo{number}{7}):\bibinfo{pages}{1--16}.
\bibitem[{Jankovic and Stojadinovic(2004)}]{Jankovic2004}
\bibinfo{author}{Jankovic\xfnm[ S.]}, \bibinfo{author}{Stojadinovic\xfnm[ B.]}.
\newblock \bibinfo{title}{{Probabilistic performance based seismic demand model
  for R/C frame buildings}}.
\newblock In: \bibinfo{booktitle}{Proc. 13th World Conf. Earthquake Eng.}
  \bibinfo{year}{2004},.
\bibitem[{Choun and Elnashai(2010)}]{Choun2010}
\bibinfo{author}{Choun\xfnm[ Y.S.]}, \bibinfo{author}{Elnashai\xfnm[ A.S.]}.
\newblock \bibinfo{title}{{A simplified framework for probabilistic earthquake
  loss estimation}}.
\newblock \bibinfo{journal}{Prob Eng Mech}
  \bibinfo{year}{2010};\bibinfo{volume}{25}(\bibinfo{number}{4}):\bibinfo{pages}{355--364}.
\bibitem[{Ghanem and Spanos(2003)}]{Ghanembook2003}
\bibinfo{author}{Ghanem\xfnm[ R.]}, \bibinfo{author}{Spanos\xfnm[ P.]}.
\newblock \bibinfo{title}{{Stochastic Finite Elements : A Spectral Approach}}.
\newblock \bibinfo{publisher}{Courier Dover Publications};
  \bibinfo{year}{2003}.
\bibitem[{Blatman and Sudret(2011)}]{BlatmanJCP2011}
\bibinfo{author}{Blatman\xfnm[ G.]}, \bibinfo{author}{Sudret\xfnm[ B.]}.
\newblock \bibinfo{title}{{Adaptive sparse polynomial chaos expansion based on
  Least Angle Regression}}.
\newblock \bibinfo{journal}{J Comput Phys}
  \bibinfo{year}{2011};\bibinfo{volume}{230}:\bibinfo{pages}{2345--2367}.
\bibitem[{Sudret et~al.(2011)Sudret, Piquard and Guyonnet}]{SudretGuyonnet2011}
\bibinfo{author}{Sudret\xfnm[ B.]}, \bibinfo{author}{Piquard\xfnm[ V.]},
  \bibinfo{author}{Guyonnet\xfnm[ C.]}.
\newblock \bibinfo{title}{{Use of polynomial chaos expansions to establish
  fragility curves in seismic risk assessment}}.
\newblock In: \bibinfo{editor}{{G. De Roeck G. Degrande}\xfnm[ G.L.]},
  \bibinfo{editor}{M\"{u}ller\xfnm[ G.]}, editors. \bibinfo{booktitle}{Proc.
  8th Int. Conf. Struct. Dynamics (EURODYN 2011), Leuven, Belgium}.
  \bibinfo{year}{2011},.
\bibitem[{Sudret and Mai(2013{\natexlab{b}})}]{SudretMaiIcossar2013}
\bibinfo{author}{Sudret\xfnm[ B.]}, \bibinfo{author}{Mai\xfnm[ C.V.]}.
\newblock \bibinfo{title}{{Computing seismic fragility curves using polynomial
  chaos expansions}}.
\newblock In: \bibinfo{editor}{Deodatis\xfnm[ G.]}, editor.
  \bibinfo{booktitle}{Proc. 11th Int. Conf. Struct. Safety and Reliability
  (ICOSSAR'2013), New York, USA}. \bibinfo{year}{2013}{\natexlab{b}},.

\end{thebibliography}







\end{document}